\documentclass[11pt]{article}%
\usepackage{amssymb}
\usepackage{authblk}
\usepackage{amsfonts}
\usepackage{amsmath}
\usepackage{graphicx}
\usepackage{amsthm}
\usepackage{color}
\usepackage{subcaption}
\usepackage{tikz}
\usepackage{epstopdf}
\usepackage{url}

\usetikzlibrary{shapes,arrows}
\tikzstyle{line} = [draw, -latex']
\providecommand{\U}[1]{\protect\rule{.1in}{.1in}}
\definecolor{mygreen}{rgb}{0.0, 0.5, 0.0}
\newcommand{\cmt}[1]{\textcolor{mygreen}{#1}} 
 
\newtheorem{theorem}{Theorem}

\begin{document}
\title{Interspecific competition shapes the structural stability of mutualistic networks}
\author[1,2,$\dagger$]{Xiangrong Wang}
\author[3,$\dagger$]{Thomas Peron}
\author[4]{Johan L. A. Dubbeldam}
\author[5,6]{Sonia K\'efi}
\author[7,8,9]{Yamir Moreno}
\affil[1]{Institute of Future Networks, Southern University of Science and Technology, People's Republic of China}
\affil[2]{Research Center of Networks and Communications, Peng Cheng Laboratory, People's Republic of China}
\affil[3]{Institute of Mathematics and Computer Science, University of São Paulo, 13566-590 São Carlos, SP, Brazil}
\affil[4]{Faculty of Electrical Engineering, Mathematics and Computer Science, P.O. Box 5031, 2600 GA Delft, The Netherlands}
\affil[5]{ISEM, CNRS, Univ. Montpellier, IRD, EPHE, Montpellier, France}
\affil[6]{Santa Fe Institute, 1399 Hyde Park Road, Santa Fe, NM 87501, USA}
\affil[7]{Institute for Biocomputation and Physics of Complex Systems (BIFI), University of Zaragoza, Zaragoza, Spain}
\affil[8]{Department of Theoretical Physics, Faculty of Sciences University of Zaragoza, Zaragoza, Spain}
\affil[9]{Institute for Scientific Interchange Foundation, Turin, Italy}
\affil[$\dagger$]{These authors contributed equally to this work.}
\date{26 January 2021}



\maketitle


\begin{abstract}

Mutualistic networks have attracted increasing attention in the ecological literature in the last decades as they play a key role in the maintenance of biodiversity. Here, we develop an analytical framework to study the structural stability of these networks including both mutualistic and competitive interactions. Analytical and numerical analyses show that the structure of the competitive network fundamentally alters the necessary conditions for species coexistence in communities. Using 50 real mutualistic networks, we show that when the relative importance of shared partners is incorporated via weighted competition, the feasibility area in the parameter space is highly correlated with May's stability criteria and can be predicted by a functional relationship between the number of species, the network connectance and the average interaction strength in the community. Our work reopens a decade-long debate about the complexity-stability relationship in ecological communities, and highlights the role of the relative structures of different interaction types.

\end{abstract}

\section{Introduction}

Species rarely live in isolation, but constantly interact with other species with
different 
interaction types, such as
predation, competition and mutualism  \cite{townsend2003essentials}. Mutualism, in which different species interact for their mutual benefit, is ubiquitous in terrestrial ecosystems~\cite{bascompte2013mutualistic}. Examples include, but are not limited to, plants receiving effective pollination or seed-dispersion by offering rewards of nutrients to their visiting animals, plants gaining resistance to insect herbivores by offering nutrients and shelter to fungi or ants, and leguminous plants obtaining nitrogen by rewarding nitrogen-fixing bacteria.

Interspecific competition, where species within the same guild compete for shared mutualistic partners, is one of the identified costs when progressing from two species mutualism to species-rich mutualism~\cite{palmer2003competition}. 
This had already been reported in field experiments of mutualistic systems of plant and pollinators by Charles Robertson \cite{robertson1895philosophy} in 1895, which was then followed by extensive studies in mutualisms of ants and plants, and of parrots and plants \cite{connell1983prevalence,campbell1985mechanism, feinsinger1987effects, davidson1981competition,palmer2003competition,stanton2003interacting,blanco2015parrots,montesinos2017network}. Intra-guild competition among plants is specifically 
recognized when common pollinators frequently visit pollen- or nectar-rich species while reducing or avoiding the visitation to less-rewarding plants \cite{mitchell2009new,aschehoug2016mechanisms,wagg2011mycorrhizal,ghazoul2006floral,fricke2017mutualistic}. Extremely attractive species can 
become dominant in the long run (e.g. well-identified representatives of \emph{Lythrum salicaria} \cite{brown2002competition} and \emph{Impatiens glandulifera} \cite{chittka2001successful}), possibly threatening the persistence of less-rewarding species.

Competition among pollinators is likewise relevant for the 
	functioning of plant-pollinator communities.
	In fact, for 
	certain hummingbird species, interspecific competition may be as 
	important as mutualistic interactions in shaping the evolution 
	of species that coexist in particular geographical 
	areas~\cite{graham2009phylogenetic,brown1985community}. More 
	specifically, as a consequence of competition, hummingbird 
	species may experience morphological changes, such as in bill 
	length, which improve the pollination efficiency or to 
	expand the diversity of pollinated 
	flowers~\cite{graham2009phylogenetic,brown1985community}. A 
	large amount of evidence suggests that morphological 
	specialization is also an evolutionary strategy to avoid or 
	reduce inter- and intra-specific competition in communities of 
	bumblebees~\cite{willmer2011pollination}. Yet the effects of 
	competition are perhaps more perceptible on shorter time scales. 
	For instance, empirical studies reveal that competition for 
	floral resources significantly alters the feeding performance 
	and the harvest of nectar in pollinator communities when 
	foreigner bees are 
	inserted 
	in a given environment~\cite{johnson1974aggression,roubik1980foraging,roubik1986sporadic}; 
	the presence of invader bees is also likely to affect the 
	availability of food and nest sites, which in turn may 
	undermine the abundance of native 
	species~\cite{fort2014quantitative,roubik1986sporadic,wilms1997floral}. 
	The immediate rearrangement of mutualistic interactions by some 
	pollinators after the intentional removal of competing 
	species is another notable example of the key role that the shared 
	use of resources plays in the dynamics of ecological communities~\cite{palmer2003competition}.

Main theories of biodiversity, however, have largely ignored the diversity of interaction types that link species in nature and have instead focused on a small subset of well studied interactions, such as predation, competition and mutualism, each of them being typically studied in isolation from the others~\cite{kefi2012more}. Decades of studies on these interactions have shown that ecological networks
have a specific architecture, which plays a key role for their dynamics and stability (e.g.~\cite{de1995energetics,neutel2002stability,bascompte2003nested,stouffer2011compartmentalization}). Mutualistic networks -- such as plant-pollinator networks, for example --- have attracted increasing attention in the ecological literature in the last decades~\cite{ings2009ecological}. These networks have been shown to be highly nested, with more specialist species interacting with a subset of the species that interact with more generalist species~\cite{bascompte2003nested},
which has been suggested to contribute to the maintenance of species diversity~\cite{thebault2010stability}.

Despite the tremendous contribution of these previous studies to the understanding of the link between the structure and dynamics of ecological networks, the lack of studies explicitly incorporating the diversity of interaction types has hindered the advancement of our understanding of the factors that drive the number of species that can coexist in a given community, one of the oldest questions in ecology~\cite{may1972will}. 

In previous studies, competitive interactions between plants sharing pollinators were modeled as an all-to-all connectivity pattern; that is, in a plant-pollinator scenario, all pollinators are considered to compete equally for plants, and all plants are assumed to compete for pollinators regardless  of the heterogeneous organization of the mutualistic interactions.
However, in real ecosystems, it seems unlikely that species would compete equally for shared partners without accounting for the structure of the mutualistic links. To which extent homogeneous competition can predict species coexistence and how variation in competition alters mechanisms that maintain biodiversity remains unknown. 
Given the empirical evidence of the structure of mutualistic networks and the lack of empirical knowledge about the structure of the associated competitive networks, it is of utmost importance to understand the effect of different competitive network structures on the community stability of mutualistic networks. 

Here, we investigate the effect of different assumptions regarding the assignment of competitive links among plants and among pollinators (and thereby the structure of the competitive networks) on the stability of ecological communities including competitive and mutualistic interactions. Real network structures were used for the mutualistic part of the ecological communities. More specifically, we investigate the ``feasible area'', i.e. the set of conditions (parameters) under which all species coexist and have a positive abundance. We develop a framework that predicts accurately the boundaries of the feasible area. 
Furthermore, we find that different competitive network structures yield significantly different feasibility patterns, showing that the structure of competitive interactions does have strong implications for the species diversity of  
multilayer networks including competition and mutualism.  

\begin{figure}[!htb]
	\centering
	\includegraphics[width=0.85\textwidth]{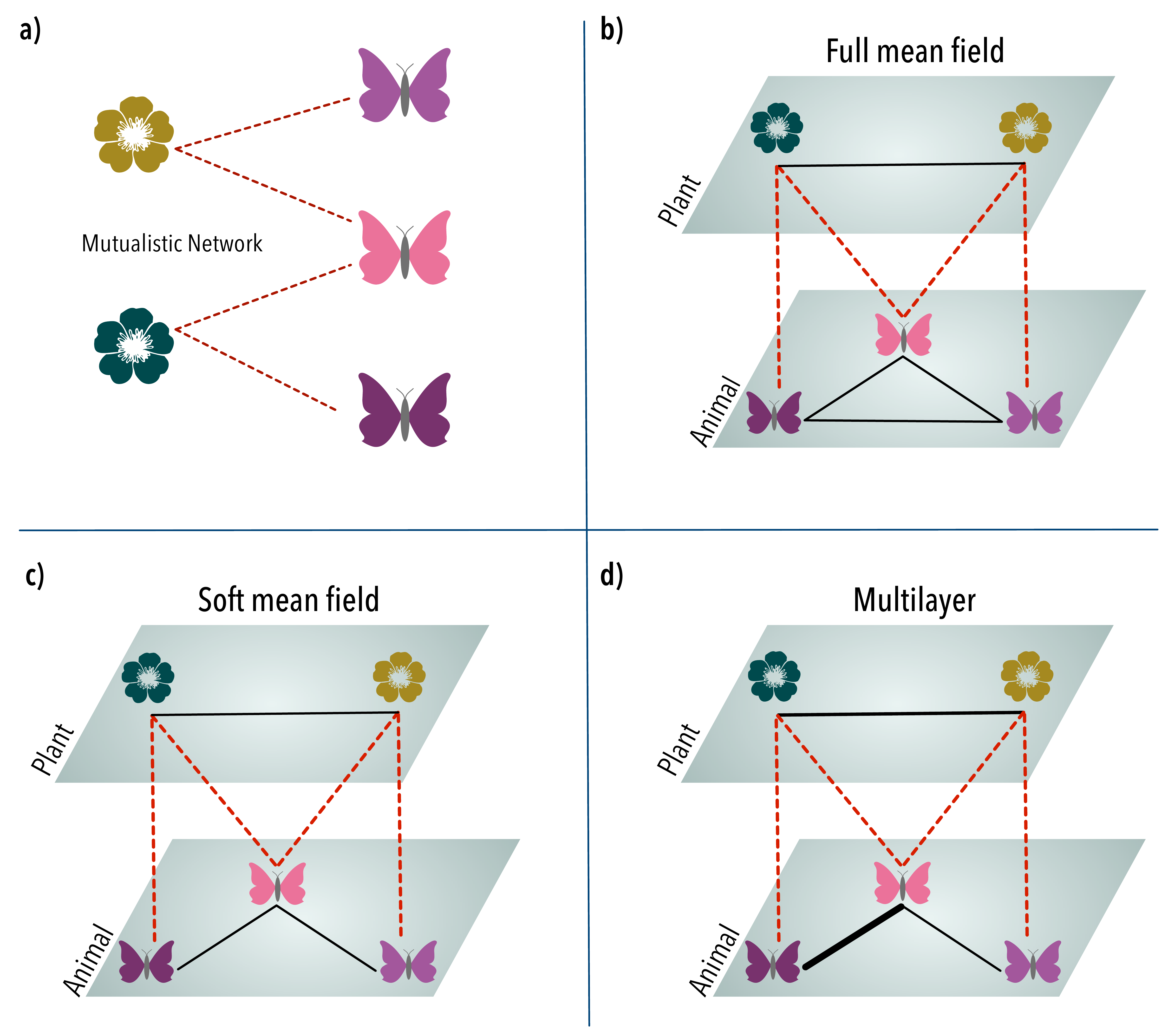}
	\caption{(a) Illustration of the minimal mutualistic network that distinguishes the (b) full mean-field, (c) soft mean-field and (d) weighted competition scenarios. For any other network with fewer nodes than shown in (a),  the competition interactions of the models in (b)-(d) become identical.  For the full mean-field, each layer is a complete, unweighted graph representing an entire inter-specific competition with the same magnitude. For the soft mean-field, each layer is an unweighted graph with connections (representing inter-specifies competition) only between pollinators (plants) who share plants (pollinators). For the weighted competition scenario, each layer is a weighted graph with weights representing the strength of inter-specific competition.}%
	\label{fig_net_toy_model_Intro}%
\end{figure}

\section{Population dynamics of competitive-mutualistic networks}

To study the impact of interspecific competition on communities persistence, we begin by describing
dynamics between plant and pollinator species. We consider a plant-pollinator system consisting of a set $ \mathcal{A} $ of $N^A$
animal species that interact mutualistically with a set $ \mathcal{P} $ of $N^P$ plant species, denoting the total biodiversity by $N = N^P + N^A$. The mutualistic interactions are fully encoded in a $N^P \times N^A$ bipartite matrix $K$, where $K_{ij}=1$ if plant species $i$ is pollinated by pollinator species $j$, and 0 otherwise. Each plant (resp. animal) species is characterized by the abundance $s_i^{P}$ (resp. $s_i^A$), whose dynamics depend on the intrinsic growth rate $\alpha_i^P$ (resp. $\alpha_i^A$) and on the
influence of competitive and mutualistic interactions as follows:   
\begin{align}
\frac{1}{s_{i}^{P}}\frac{ds_{i}^{P}}{dt} & =\alpha_{i}^{P}-\beta s_{i}^{P}-\beta_{0}\sum_{j\neq i}^{N^{P}}s_{j}^{P}+\gamma_{0}\frac{M_{i}^{P}}{1+h\gamma_{0}M_{i}^{P}}\;\;\;\;\;\;\;\;\textrm{ (Full mean-field competition), }\label{eq:fullmf_model}\\
\frac{1}{s_{i}^{P}}\frac{ds_{i}^{P}}{dt} & =\alpha_{i}^{P}-\beta s_{i}^{P}-\beta_{0}\sum_{j=1}^{N^{P}}A_{ij}^{P}s_{j}^{P}+\gamma_{0}\frac{M_{i}^{P}}{1+h\gamma_{0}M_{i}^{P}}\;\;\;\textrm{ (Soft mean-field competition),}\label{eq:softmf_model}\\
\frac{1}{s_{i}^{P}}\frac{ds_{i}^{P}}{dt} & =\alpha_{i}^{P}-\beta s_{i}^{P}-\beta_{0}\sum_{j=1}^{N^{P}}\left(\frac{W_{ij}^{P}}{M_{i}^{P}}\right)s_{j}^{P}+\gamma_{0}\frac{M_{i}^{P}}{1+h\gamma_{0}M_{i}^{P}}\textrm{ (Weighted competition),\label{eq:weighted_model}}
\end{align}
where $i=1,...,N^P$, and $M_i^P = \sum_{k \in \mathcal{A}} K_{ik} s_k^A$ is the total abundance of pollinators interacting with plant $i$. The first term in the right-hand side of the above equations corresponds 
to the intrinsic growth of each species; the second and fourth are also identical in all models and correspond to the intra-species competition and mutualistic interactions, respectively; and $ \beta $ refers to the intensity of intra-specific competition. The intensities of inter-specific competition and mutualism are denoted by $ \beta_0 $ and $ \gamma_0 $, respectively. Parameter $ h $, known as the handling time, imposes a nonlinear saturation effect on mutualism.
What distinguishes the models in Eqs.~(\ref{eq:fullmf_model})-(\ref{eq:weighted_model}) is the definition of the third term, which accounts for the intra-guild competition. We highlight schematically the differences between each competition scenario with Fig.~\ref{fig_net_toy_model_Intro}. In the ``full mean field'' model, all species within a layer compete equally with each other, i.e., all plants compete with all other plants with the same intensity [see Fig.~\ref{fig_net_toy_model_Intro}(a)], and all pollinators compete with other pollinators with the same intensity, irrespective of the mutualistic links existing between plants and pollinators (this is the approach used in previous studies; see, e.g., \cite{bastolla2009architecture}). 
In the ``soft mean-field'' model, the homogeneous competition assumption is relaxed by placing a competitive link between plants $i$ and $j$ ($A_{ij}^P=1$) only if they share at least one pollinator. The intra-guild competition links in this model are encoded in matrices $A^P$ and $A^A$. Notice that, in this scenario, the competitive links are not weighted (only present or absent). 
Finally, in the ``weighted competition'' model, the intra-guild competitive links have the same structure as in the ``soft mean-field'' model but are now weighted by the abundance of the shared partner [see Fig.~\ref{fig_net_toy_model_Intro}(c)]. This weight is set via matrix $W^P$, whose elements are given by $W_{ij}^P = \sum_{k \in \mathcal{A}} K_{ik} K_{jk}s_k^A$ [equations for the pollinator abundances $s^A_i$ follow \textit{mutatis mutandis} from Eqs.~(\ref{eq:fullmf_model})-(\ref{eq:weighted_model})]. Therefore, in the weighted scenario,  
the higher the abundance of mutualistic partners, the stronger the competition among plants which have common mutualistic connections. Notice also that the intra-guild competition terms in Eq.~(\ref{eq:weighted_model}) are asymmetric, since the biomass of shared species is normalized by the total biomass of mutualistic partners $\left(\sum_{j \in \mathcal{P}, i\neq j} W_{ij}^P/M_i^P \right)$. In other words, two plant species $i$ and $j$ perceive the competition with one another differently according to the importance of their shared pollinators in relation with their respective total abundance of pollinators.

In the Supplemental Material, we provide an exact and thorough bifurcation analysis of the toy network depicted in Fig.~\ref{fig_net_toy_model_Intro}. 
In the next section we present an analytical calculation of the solutions of Eqs.~(\ref{eq:fullmf_model})-(\ref{eq:weighted_model}) for arbitrary networks.  

\section{Structural stability conditions}

Our goal here is to derive an analytical calculation for the feasible equilibrium solution of the nonlinear population dynamics in Eqs.~(\ref{eq:fullmf_model})-(\ref{eq:weighted_model}), i.e., the solution that corresponds to the maximum biodiversity ($s_i^{P,A}>0$ $\forall i$). It is argued \cite{rohr2014structural} that a specific parameterization can be inconclusive for empirical networks due to the strong dependence of species coexistence on parameterization. Accordingly, we investigate a range of parameter values termed feasible area under which all species coexist. Henceforth, we refer to the ``feasible area'' as the region in the space spanned by parameters $\beta_0$ and $\gamma_0$ in which all species have positive abundances at equilibrium.

\subsubsection*{Condition for the full and soft mean field competitions}

We first address the solution for the full and soft mean-field models. By applying a linear approximation to the nonlinear mutualism term in Eqs.~(\ref{eq:fullmf_model}) and~(\ref{eq:softmf_model}), we rewrite the population dynamics  in a matrix form as
\begin{equation}\label{key}
\begin{bmatrix}
\frac{ds^P}{dt}\\
\frac{ds^A}{dt}
\end{bmatrix} = \text{diag}\left(\begin{bmatrix}
s^P \\
s^A
\end{bmatrix}\right)\left(\begin{bmatrix}
\alpha^P\\
\alpha^A
\end{bmatrix}-\left(\beta I + \beta_0
\begin{bmatrix}
A^P & 0\\
0 & A^A
\end{bmatrix}
-
\gamma_0\begin{bmatrix}
0& \cmt{\widetilde{M}^P}\\
\widetilde{M}^A & 0
\end{bmatrix}
\right)
\begin{bmatrix}
s^P \\
s^A
\end{bmatrix}\right),
\end{equation}
where $A^{P,A}$ and $\widetilde{M}^{P,A}$ are the matrices that set the competition and mutualism interactions, respectively; and $s^{P,A}$ are the vectors containing the individual abundances of plant and pollinators, respectively. In the full mean-field model, we have $A^P_{ij} = 1$ for $i \neq j$, while in the soft mean-field model $A^P_{ij}=\Theta\left(\sum_{k \in \mathcal{A}} K_{ik} K_{jk}\right)$, where $\Theta (\cdot)$ is the heaviside function. The elements of matrices $\widetilde{M}^{P,A}$
are obtained by applying the Taylor expansion for the mutualistic term in Eqs.~(\ref{eq:fullmf_model})-(\ref{eq:softmf_model}), that is: 
\begin{equation}\label{eq_linear_approx_nonlinear_mutualism}
\frac{\gamma_0 M_i}{1+h\gamma_0M_i} = \frac{\gamma_0 \left(M_0\right)_i}{1+h\gamma_0\left(M_0\right)_i} + \left(\frac{\gamma_0 M_i}{1+h\gamma_0M_i}\right)^\prime\Bigg|_{M_i = \left(M_0\right)_i}(M_i - \left(M_0\right)_i),
\end{equation}
expanded around a point $(M_0)_i$ 
close to a fixed point that is challenging to obtain without a prior knowledge. We sometimes omit the subscript $(M_0)_i$ when there is no ambiguity. After substituting both the competitive and mutualistic terms, a feasible equilibrium ($ s_i^P, s_i^A > 0 $) can thus be obtained by solving the following linear equation
\begin{equation}\label{eq_feasibbe_beta_gamma_approx_meanField}
\begin{bmatrix}
\alpha^P\\
\alpha^A
\end{bmatrix}=\left(\beta I + \beta_0
\begin{bmatrix}
A^P &0\\
0& A^A
\end{bmatrix}
-
\begin{bmatrix}
0& \text{diag}\left(\frac{\gamma_0}{\left(1+h\gamma_0\left(M_0^P\right)_i\right)^2}\right)K\\
\text{diag}\left(\frac{\gamma_0}{\left(1+h\gamma_0\left(M_0^A\right)_i\rangle\right)^2}\right)K^T& 0
\end{bmatrix}
\right)
\begin{bmatrix}
s^P \\
s^A
\end{bmatrix} - c,
\end{equation}
where the vector  $ c = h\left(\left(\frac{\gamma_0M_0^P}{1+h\gamma_0M_0^P}\right)^2,\ \left(\frac{\gamma_0M_0^A}{1+h\gamma_0M_0^A}\right)^2 \right)^T$. Without a prior knowledge on the fixed points of the system, the linearizion of the system near a fixed point appears to be challenging or even unfeasible. We approach this challenge by analyzing the interplay between the mutualistic interactions and the intra-guild competition, which separately lead to abundance gain and abundance loss at equilibrium. When the mutualistic strength is equal to the competition strength, the species abundance on average follows $ \langle s_i \rangle = \frac{\alpha_i}{\beta_i} $. Assuming the average abundance $ \langle s_k^A \rangle = \langle s_i \rangle $ for all the animal species pollinating plant $ i $, we linearize the nonlinear population dynamics at $ \left(M_0^P\right)_i = d_i^P \langle s_i \rangle $ for each plant $ i $, where $ d_i^P = \sum_{k}K_{ik} $ denoting the number of animals pollinating plant $ i $. The fixed point for animal species $ M_0^A $ is approximated similarly.

Equation~(\ref{eq_feasibbe_beta_gamma_approx_meanField}) provides an approximated solution for the abundances of general networks in the full and soft-mean field competition. Notice that Eq.~(\ref{eq_feasibbe_beta_gamma_approx_meanField}) provides the equilibrium solution of the system, but does not guarantee feasibility. In order to estimate the feasible area, one needs to solve Eq.~(\ref{eq_feasibbe_beta_gamma_approx_meanField}) 
for different parameters seeking solutions satisfying $s_i^{P,A} > 0$.  Equation~\ref{eq_feasibbe_beta_gamma_approx_meanField} has the numerical advantage that it allows one to scan the parameter space of ecological networks and delineate the feasible area much more quickly than
by evolving the original dynamics. Differently from the solutions assuming $ h = 0  $ in \cite{rohr2014structural}, Equation~(\ref{eq_feasibbe_beta_gamma_approx_meanField}) is applicable to any real $ h \geq 0 $, thus enabling the investigation of various mutualistic regimes.

\begin{figure}[!tb]
	\centering
	\includegraphics[width=1.0\textwidth]{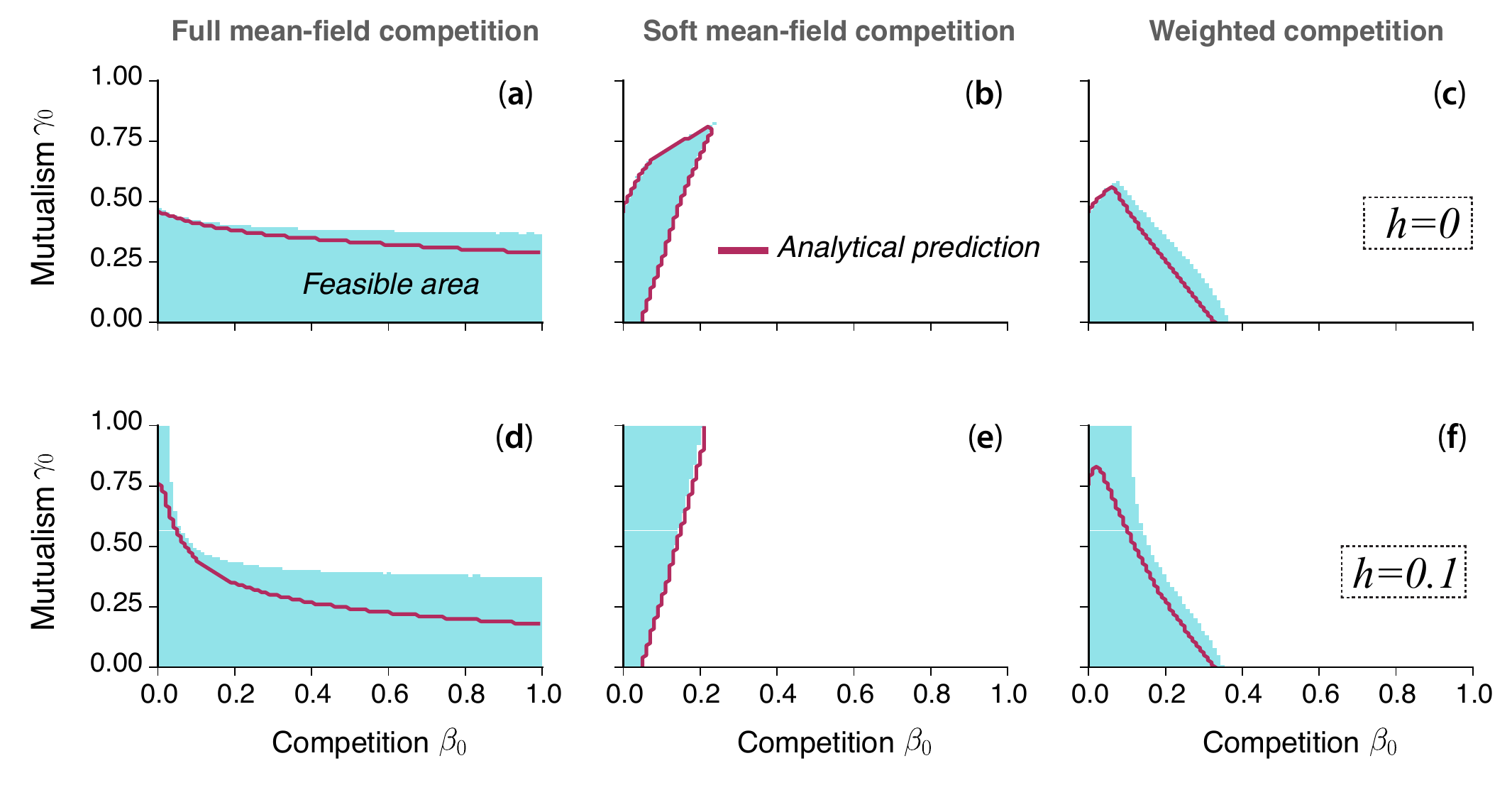}
	\caption{Feasible area patterns illustrated here for a real plant-pollinator network (MPL-16) from the Web of Life platform~\cite{weboflife}, in the (left) full mean-field, (center) soft mean-field and (right) weighted scenarios. A point $(\beta_0,\gamma_0)$ is colored in blue if all species survive with a positive abundance in the stationary regime of the simulations of Eqs.~(\ref{eq:fullmf_model})-(\ref{eq:weighted_model}) for that parameter choice. Parameters: $\alpha_i=1\forall i$, and $\beta=5$. Diagrams in the upper panels have $h=0$, while simulations in the lower panels are for $h=0.1$. Grid size: $100\times 100$. Solid lines are obtained by solving Eq.~(\ref{eq_feasibbe_beta_gamma_approx_meanField}).}
	
	\label{fig1}
\end{figure}

In Fig.~\ref{fig1}(a,d) and (b,e) we compare the analytical predictions provided by Eq.~(\ref{eq_feasibbe_beta_gamma_approx_meanField}) with direct simulations of the systems in Eqs.~(\ref{eq:fullmf_model})-(\ref{eq:weighted_model}) over the parameter space spanned by competition and mutualism strengths, $\beta_0$ and $\gamma_0$, respectively. 
As it is seen, the analytical prediction delineates the boundaries of the feasible area with remarkable accuracy 
for the soft mean-field model, for both $h=0$ and $h=0.1$. In the full mean-field model, reasonable precision is achieved for $h=0$, while for $h=0.1$ the matching between numerical and theoretical boundaries is lost as competition strength increases. In the Supplemental Material, we show that the approximate solution of Eq.~(\ref{eq_feasibbe_beta_gamma_approx_meanField}) is successful in predicting the feasible area for several real networks under the full mean-field and soft mean-field competition scenarios.  

\subsubsection*{Condition for the weighted competition}
To analyze the impact of heterogeneity in the competitive strengths among intra-guild species, we derive conditions under which species coexist for the dynamical model with weighted competition. The weighted competition is reduced to the homogeneous competitive strength (soft-mean field competition) only when any pair of intra-guild species depend on and share exactly the same set of mutualistic partners, corresponding to complete-like bipartite mutualistic networks.

In the dynamic model for weighted competition [Eq.~(\ref{eq:weighted_model})], the inter-specific competition interactions are weighted by the relative importance of shared resources in a nonlinear form. To tackle the nonlinear inter-specific competition, we harness the microscopic perspective of intra-guild competition induced by a single mutualistic plant-pollinator interaction. When plant $ i $ is pollinated by an animal $ k $ (i.e., $ K_{ik} = 1 $), the inter-specific competition between plant $ i $ and the other plants $ j $ that are pollinated by animal $ k $ reads $ \sum_{j \in \mathcal{P}, j \neq i}K_{ik}K^T_{kj}s_j^P $. Summing over all the pollinators $ k $ that pollinate plant $ i $ yields the total inter-specific competition $ \sum_{k} \left(\sum_{j \in \mathcal{P}, j \neq i} K_{ik}K^T_{kj}s_j^P\right)s_k^A $ for that plant $ i $. Armed with the view of a single mutualistic interaction, the nonlinear competition term in Eq. (\ref{eq:weighted_model}) can be rewritten as
\begin{equation}\label{eq:weighted_competition_term}
\sum_{j \in \mathcal{P}, j \neq i} \frac{W_{ij}^P}{M_i^P}s_j^P = \frac{\sum_{k} \left(\sum_{j \in \mathcal{P}, j \neq i} K_{ik}K^T_{kj}s_j^P\right)s_k^A}{\sum_{k}K_{ik}s_k^A}
\end{equation}
Analytical estimates can be obtained by reigning in the weighted competition terms. This can be accomplished by using the mediant inequality (see SI material):
\begin{equation}\label{eq_inequality_intra_competition}
\min_{k}\sum_{j \in \mathcal{P}, j \neq i} K_{ik}K^T_{kj}s_j^P \leq \frac{\sum_{k} \left(\sum_{j \in \mathcal{P}, j \neq i} K_{ik}K^T_{kj}s_j^P\right)s_k^A}{\sum_{k}K_{ik}s_k^A} \leq \max_{k}\sum_{j \in \mathcal{P}, j \neq i} K_{ik}K^T_{kj}s_j^P
\end{equation}

Equation~(\ref{eq_inequality_intra_competition}) allows reducing the complexity of species abundance from two-guilds into a single-guild species abundance. After the complexity reduction, the weighted competition structure can be encoded in a competition matrix $ \tilde{A}^P  $ (resp. $ \tilde{A}^A $) for plant (resp. animal) species, in analogy to the competition matrix, incorporated in the unweighted adjacency matrix $ A^P $, for the soft mean-field case. Additionally, Eq.~(\ref{eq_inequality_intra_competition}) establishes lower and upper bounds for the competition term.

Seeking to find an accurate estimate for the competition term, it is appropriate to consider that plant $ i $ competes with plant $ j $ mediated by sharing the animal $k$, whose degree is, over all pollinators of plant $i$, the closest to the local average of pollinated plants. Specifically, the element of the competition matrix $ \tilde{A}^P $ can be written as
\begin{equation}
\tilde{A}_{ij}^{P}=\begin{cases}
1 & \textrm{ if }K_{ik}K_{kj}^{T}\neq0\textrm{ for }k\in\arg\min_{k\in A}\left|d_{k}^{A}-\left\lfloor \frac{\sum_{s}K_{is}d_{s}^{A}}{d_{i}^{P}}\right\rfloor \right|,\\
0 & \textrm{otherwise,}
\end{cases}
\label{eq:approx_weighted_scenario}
\end{equation}
where  $d_i^P$ ($d_i^A$) is the number of animals (plants) with which plant (animal) $i$ interacts, meaning that the weighted competition is approximated by an \emph{effective} mutualistic partner $ k $ whose degree is the closest to the average competitive species per mutualistic interaction.

Combining Eq.~(\ref{eq:approx_weighted_scenario}) with the corresponding mutualistic term into an expression analogous to Eq. (\ref{eq_feasibbe_beta_gamma_approx_meanField}), the feasible solution for weighted competition model is eventually obtained by
\begin{equation}\label{eq_feasibbe_beta_gamma_approx_multilayer}
\begin{bmatrix}
s^P \\
s^A
\end{bmatrix} =\left(\beta I + \beta_0
\begin{bmatrix}
\tilde{A}^P &0\\
0& \tilde{A}^A
\end{bmatrix}
-
\begin{bmatrix}
0& \text{diag}\left(\frac{\gamma_0}{\left(1+h\gamma_0\left(M_0^P\right)_i\right)^2}\right)K\\
\text{diag}\left(\frac{\gamma_0}{\left(1+h\gamma_0\left(M_0^A\right)_i\rangle\right)^2}\right)K^T& 0
\end{bmatrix}
\right)^{-1}
\left(\begin{bmatrix}
\alpha^P\\
\alpha^A
\end{bmatrix}+c\right)
\end{equation}

The analytical prediction of Eq.~(\ref{eq_feasibbe_beta_gamma_approx_multilayer}) for the weighted model is also well confirmed by numerical simulations on real mutualistic networks 
[Fig.~\ref{fig1}(c) and (f)]. Besides checking the validity of our calculations, Fig.~\ref{fig1} also allows us to highlight the marked differences in the dynamics yielded by the three competition models. Notice, in particular, how strongly the shape of the feasible area changes from the full mean-field to the soft mean-field and then to the weighted competition case: the mere inclusion of heterogeneity in the competition term in Eq.~(\ref{eq:fullmf_model}) shifts the region of occurrence of feasible states from strong competition (full mean-field) to weak competition (soft mean-field, weighted). Another noteworthy difference between soft-mean field and weighted competition is that the former allows a wider feasible region for low competition and high mutualism, while the latter is favored by modest values of both competition and mutualism. These results clearly demonstrate how crucial the structure of the intra-guild competition networks is for the species coexistence and the diversity of ecological communities.

\section{Stability-complexity paradox}

After investigating the impact of different choices of the structure of intra-guild competition networks on biodiversity, let us now study how the feasible area relates with the topological properties of the multilayer ecological networks. 
To do so, we consider a set of 50 real plant-pollinator networks retrieved from the Web of Life platform~\cite{weboflife} and, for each network in the database, we evolve Eqs.~(\ref{eq:fullmf_model})-(\ref{eq:weighted_model}) and calculate the corresponding feasible area for each intra-guild competition scenario.

\begin{figure}[!tb]
	\centering
	\includegraphics[width=1.0\textwidth]{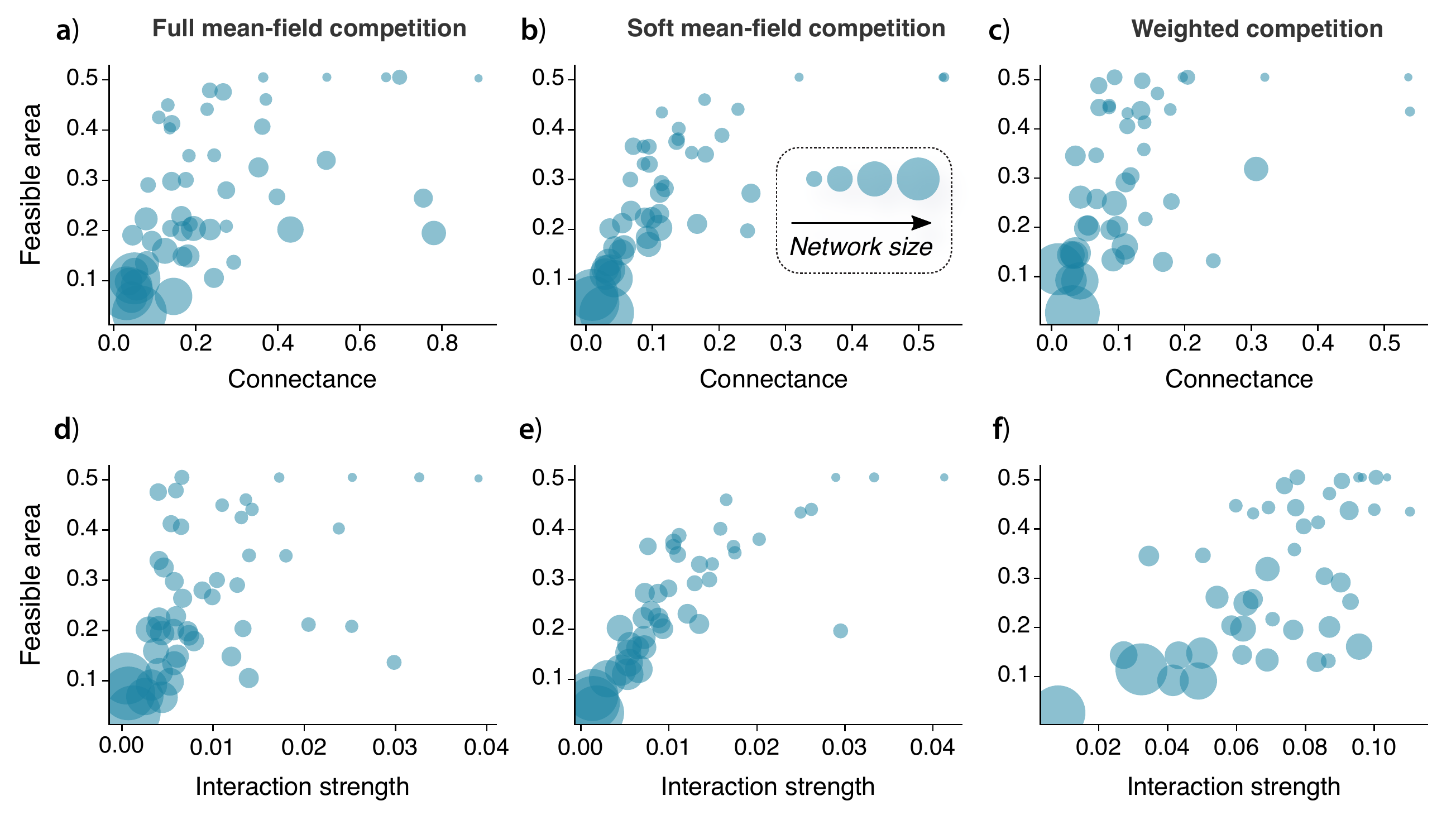}
	\caption{
	Feasible area as a function of different network metrics for full mean field [(a) and (d)], soft mean-field [(b) and (e)], and weighted [(c) and (f)] competition scenarios. 
	Each dot corresponds to a real mutualistic network from the Web of Life platform~\cite{weboflife}, with the size of dot proportional to the network size. Parameters common in all panels: $\alpha_i=1\forall i$, $h=0.1$ and $\beta=5$. Feasible area was calculated considering the same ranges of $\beta_0$ and $\gamma_0$ shown in Fig.~\ref{fig1}.
	}%
	\label{fig2}%
\end{figure}

We look at different components of complexity independently, namely species diversity, the average interaction strength and connectance. 
Following May~\cite{may1972will}, we define the average interaction strength as the average of the off-diagonal elements of the Jacobian matrices of systems (\ref{eq:fullmf_model})-(\ref{eq:weighted_model}); and connectance is defined as the density of non-zero values of the Jacobian matrices (see Supplemental Material). 
As shown in Fig.~\ref{fig2}, the feasible area correlates positively with connectance and interaction strength in  
the full mean-field, soft mean-field and weighted competition scenarios. In other words, the higher the number of interactions among species and the stronger their intensity, the more likely that the ecological community exhibits feasible states. This result actually points back to the long debated diversity-stability paradox initiated by May~\cite{may1972will}. In his work, May proved that the probability of facing stable states converges almost certainly to zero for sufficiently large communities. In mathematical terms, suppose that species $i$ and $j$ 
interact with probability $C$ and via an interaction strength $J_{ij}$, which is a random variable with mean $\mathbb{E}(J_{ij})=0$ and with variance given by $\mbox{Var}(J_{ij})=\sigma^2$. Under such conditions, and setting the self-interaction terms as constants, $J_{ii}=-d \;\forall i$, May proved that the dynamical system $d \mathbf{s}/dt = \mathbf{J}\mathbf{s}$ is almost surely stable if  
\begin{equation}
 \sqrt{NC} < \frac{d}{\sigma}, 
\label{eq:may_cond_trad}
\end{equation}
in particular with $ d=1 $, i.e. the condition for which the leading eigenvalue of $\mathbf{J}$ is negative. Consequently, the increases in size, connectivity and interaction strength favor the dynamical destabilization of the system. This finding triggered the aforementioned paradox because it seems in contradiction with the high diversity of species observed in many natural communities~\cite{mccann2000diversity}.
 
At first sight, our results seem to violate the stability-diversity paradox, since they show a positive correlation between feasible area and connectance, and between feasible area and interaction strength,  suggesting that more connected --and thereby more complex-- networks tend to have a higher feasibility area, meaning that such systems would tend to be more stable  
[Fig.~\ref{fig2}, panels (a) and (c)]. However, this conclusion is reached by looking at the different aspects of complexity independently, whereas a closer inspection suggests that some of these aspects are related.
For example, incorporating the third element of May's relation and inspecting the size of the networks,
we realize that the networks with a high connectance are also those that are the smallest networks in the database (Fig.~\ref{fig2}, size of the dots stands for number of species in the corresponding networks). What is more, by checking manually the networks with connectance values $\gtrsim0.12$, one notices that they correspond to almost fully connected bipartite matrices and, hence, to almost fully connected Jacobian matrices as well. 
Thus, the networks with the highest possible values for the feasible area in Fig.~\ref{fig2} are in fact the networks with the less ``complex'' structure in the data set, in consonance with the notion that complexity, as quantified by this trade-off between system size and connectance, tends to destabilize ecological communities. 

\begin{figure}[!tb]
	\centering
	\includegraphics[width=1.0\textwidth]{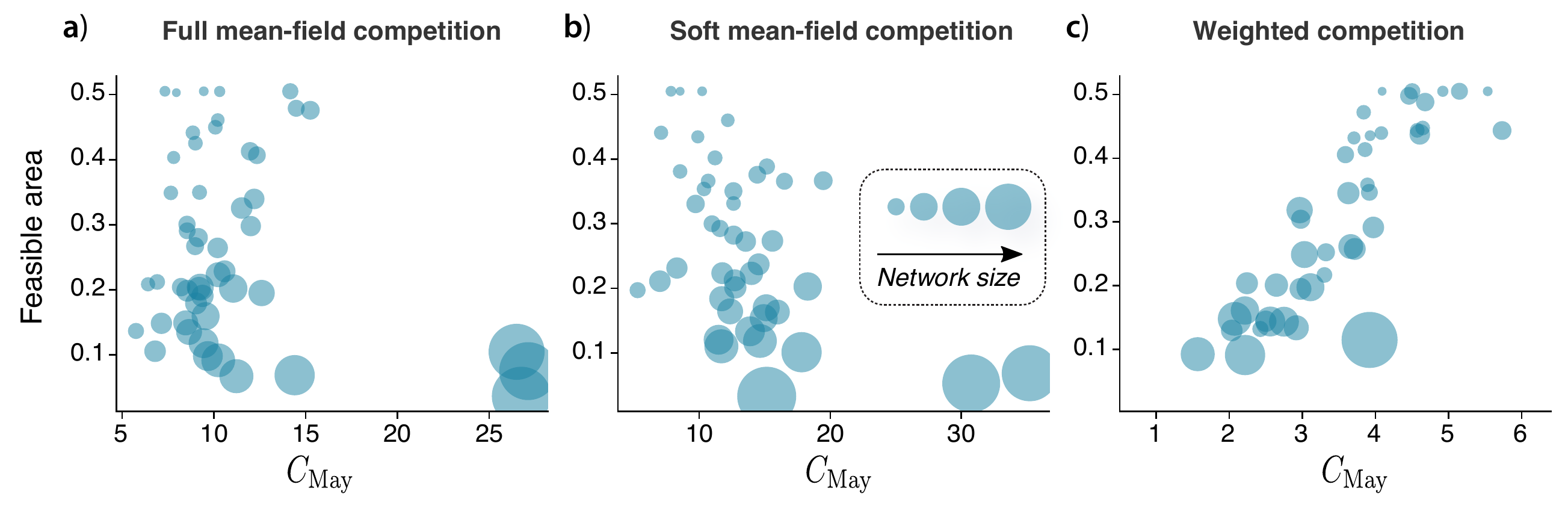}
	\caption{Comparison between feasible area and May's stability condition [Eq.~(\ref{eq:conditions_betagamma})] for several real plant-pollinator networks considering (a) full mean-field, (b) soft mean-field and (c) weighted intra-guild competition schemes. Parameters common in all panels: $\alpha_i=1\forall i$, $h=0.1$ and $\beta=5$.}
	\label{fig3}%
\end{figure}

We now look more specifically at May's criteria combining the three network metrics and explore how the size of the feasibility area is related to that criteria. Our goal here is to check whether there is a clear relation between the feasible area and the expressions in Eqs.~(\ref{eq:fullmf_model})-(\ref{eq:weighted_model}). However, it is noteworthy that May's condition [Eq.~(\ref{eq:may_cond_trad})] is related to the probability that the ecological system is stable for a particular set of parameters, whereas the feasible area results from a sum over different parameter combinations. Therefore, to appropriately evaluate how the dependencies of the feasible area on network properties relate with Eq.~(\ref{eq:may_cond_trad}), we define the following quantity:
\begin{equation}
C_{\mbox{\small{May}}}=\langle\langle J_{ii}\rangle\rangle_{(\beta,\gamma)}-\left\langle \sigma(J_{ij})\right\rangle _{(\beta,\gamma)}\sqrt{NC},
\label{eq:conditions_betagamma}
\end{equation}
where $\langle \cdot \rangle$ corresponds to an average over the Jacobian matrix's elements,
and  $\langle  \cdot  \rangle_{(\beta,\gamma)}$ stands for the average over the parameters $\beta$ and $\gamma$ considered in Fig.~\ref{fig2}. The first term in Eq.~(\ref{eq:conditions_betagamma}), $\langle\langle J_{ii} \rangle \rangle_{(\beta,\gamma)}$, is the average taken over the diagonal elements, since, contrarily to the random model considered by May, the diagonal elements of the Jacobian, $J_{ii}^{\rm{P,A}}$, are not constant, but rather are heterogeneously distributed over the diagonal (see Supplemental Material); the term $\langle \sigma(J_{ij})  \rangle_{(\beta,\gamma)}$ is the average standard deviation of the off-diagonal values of $\mathbf{J}$. In practical terms, variable $C_{\textrm{May}}$ defined in Eq.~(\ref{eq:conditions_betagamma}) quantifies how distant a given network is from the critical point established by May's stability condition, meaning how stable it is according to that criteria. For each real network considered, we numerically calculate the Jacobian elements evaluated at the stationary points, which in turn are obtained by evolving Eqs~(\ref{eq:fullmf_model})-(\ref{eq:weighted_model}) numerically. We visualize the dependence of the  feasible area on $C_{\textrm{May}}$ in Fig.~\ref{fig3}. 
Interestingly, the size of the feasibility area does not correlate with May's criteria for the full and soft mean field scenarios, but it is very strongly correlated to it in the case of the weighted competition scenario, 
despite the fact that May's criteria was formulated for much more idealized systems (Fig.~\ref{fig3}). The patterns we see in Fig.~\ref{fig3} also agree with the scatter plots in Fig.~\ref{fig2}; i.e., there is a noticeable correlation between network size and the value of the coefficient $C_{\textrm{May}}$. In particular, we observe that large networks tend to exhibit low values for $C_{\textrm{May}}$, while ``less complex'' networks (in terms of size and connectivity) have higher values of $C_{\textrm{May}}$.

This raises the question of why the weighted competition model adheres with May's criteria better than the full and soft mean-field scenarios. The answer lies in the expression of the Jacobian terms of the different models
in Eqs.~(\ref{eq:fullmf_model})-(\ref{eq:weighted_model}). For the weighted competition scenario, the off-diagonal Jacobian elements $J_{ij}$ are proportional to the terms $1/s_{i}^{P,A}$ and $1/(s_{i}^{P,A})^2$ (see the SI material). This makes the elements $J_{ij}$ to narrowly peak around an average value, thus making the standard deviation $\sigma(J_{ij})$ less significant 
than the average of the diagonal terms $J_{ii}$ (see SI material) and thereby creating a positive correlation between feasible area and $C_{\rm{May}}$. 
Comparing the models of heterogeneous competitive interactions (i.e. soft mean-field and weighted in Fig.~\ref{fig3}), one notices that the introduction of weights in the competitive structure acts as a stabilizing factor in the dynamics, in the sense that networks tend to exhibit a wider feasible area in the weighted scenario. 
The interplay of interaction types among species, a well-defined structure, and interacting strength shows a markedly driving force in stabilizing the system that is highly predictable and consistent with May's criterion.
Interestingly, the feasible area of the weighted competition scenario correlates with May's criterion as initially formulated by May for random matrices, but no significant relation was observed with other, more recent stability criteria formulated for random matrix models that incorporate features from predator-prey and mutualistic interactions~\cite{allesina2012stability} (see SI material).

\section{Discussion}

We investigated to which extent the incorporation of intra-guild competition alters the maintenance of biodiversity in mutualistic systems. Compared to a scenario where all species from a guild homogeneously compete with each other, as commonly assumed in the literature, heterogeneous competition leads to markedly different patterns for the feasibility area of plant-pollinator networks. 
Without sufficient empirical data about intra-guild competition in plant-pollinator communities, deriving 
the structure of the competitive links 
from the observed mutualistic interactions enables us to theoretically explore how the structure of different intra-guild competition networks affects the structural stability of mutualistic ecological communities.

Our results show that previously identified implications can be restricted to homogeneous competition and cannot be readily generalized to heterogeneous competition. Specifically, we found that feasibility patterns are dramatically modified when competitive interactions become heterogeneous. This finding suggests that a series of important conclusions regarding the dynamics of ecological networks might have been overlooked given that theoretical models have been traditionally studied under the assumption of homogeneous intra-guild competitive interactions. Therefore, getting information on the structure of competitive networks in mutualistic systems is key to better understand what constrains the assembly of mutualistic communities during the dynamical coevolutionary process, which could be an important driving force of coevolution~\cite{guimaraes2017indirect}. 

Finally, by investigating the feasible area in terms of global network properties, we found that smaller and less connected networks exhibited larger regions sustaining maximum biodiversity. Interestingly, this result agrees with the long-standing May's stability-diversity paradox, which states that complex systems are more prone to be destabilized as their size, connectance and mean interaction strength increase. Indeed, we have verified that the more structurally stable networks turned out to be the ``less complex'' ones according to May's criterion~\cite{may1972will}. Our results therefore show that the complexity introduced in the model by the weighted competition scenario yields a phenomenology which is well predicted by a condition originally derived for random systems [Fig.~\ref{fig3}(c)], whereas it is not the case for the other competition scenarios. Importantly, the analysis on feasible solutions performed here is not limited to the specific equations studied here, but can also be extended to ecological networks with different types of interactions, such as facilitation~\cite{zele2018ecology}, and predator-prey models~\cite{gross2009generalized}. 

\section*{Acknowledgments}
X.W. acknowledges the project 62003156 supported by NSFC and ``PCL Future Greater-Bay Area
Network Facilities for Large-scale Experiments and Applications (LZC0019)''. T.P. acknowledges FAPESP (Grants No. 2016/23827-6 and No. 2018/15589-3). This research was carried out using the computational resources of the Center for Mathematical Sciences Applied to Industry (CeMEAI) funded by FAPESP
(Grant No. 2013/07375-0). Y.M. acknowledges partial support from the Government of Aragon and FEDER funds, Spain through grant E36-20R (FENOL), by MINECO and FEDER funds (FIS2017-87519-P), and by Intesa Sanpaolo Innovation Center. The funders had no role in study design, data collection, and analysis, decision to publish, or preparation of the manuscript. 


\newpage

\section*{Supplementary Material}

\section{Exact solution for the structural stability of the minimal model}

For the sake of clarity, let us briefly reintroduce the models discussed in the main text. We consider plant-pollinator systems composed of $N^A$ animal species, which interact with $N^P$ plant species. The matrix encoding the mutualistic is the $N^P \times N^A$ bipartite matrix $K$, whose elements are $K_{ij}=1$ if plant species $i$ is pollinated by animal species $j$. The abundance of the $i$th plant is defined as $s_i^P$ ($s_i^A$, analogously for the animal species), and its time-depend dynamics is governed by the following equations:
\begin{align}
\frac{1}{s_{i}^{P}}\frac{ds_{i}^{P}}{dt} & =\alpha_{i}^{P}-\beta s_{i}^{P}-\beta_{0}\sum_{j\neq i}^{N^{P}}s_{j}^{P}+\gamma_{0}\frac{M_{i}^{P}}{1+h\gamma_{0}M_{i}^{P}}\;\;\;\;\;\;\;\;\textrm{ (Full mean-field competition), }\label{eq:fullmf_model}\\
\frac{1}{s_{i}^{P}}\frac{ds_{i}^{P}}{dt} & =\alpha_{i}^{P}-\beta s_{i}^{P}-\beta_{0}\sum_{j=1}^{N^{P}}A_{ij}^{P}s_{j}^{P}+\gamma_{0}\frac{M_{i}^{P}}{1+h\gamma_{0}M_{i}^{P}}\;\;\;\textrm{ (Soft mean-field competition),}\label{eq:softmf_model}\\
\frac{1}{s_{i}^{P}}\frac{ds_{i}^{P}}{dt} & =\alpha_{i}^{P}-\beta s_{i}^{P}-\beta_{0}\sum_{j=1}^{N^{P}}\left(\frac{W_{ij}^{P}}{M_{i}^{P}}\right)s_{j}^{P}+\gamma_{0}\frac{M_{i}^{P}}{1+h\gamma_{0}M_{i}^{P}}\textrm{ (Weighted competition),\label{eq:weighted_model}}
\end{align}
where $\alpha^{P,A}_i$ are the intrinsic growth rate; parameters $\beta$ and $\beta_0$ stand for the intra- and inter-specific competition strength, respectively; and $\gamma_0$ is the mutualism strength. Variable $M_i^P$ is given by $M_i^P = \sum_{k\in A} K_{ik}s_k^A$, where $k$ denotes the indexes belonging to the pollinators set. Matrix $A_{ij}^P$ of the ``Soft mean-field competition'' scenario encodes the competitive connections within the plant-guild: $A_{ij}^P=1$ if plants $i$ and $j$ share at least one pollinator, and 0 otherwise. For the ``Weighted competition scenario'', the elements of matrix $W^P$ are defined as $W_{ij}^P = \sum_{k \in A} K_{ik} K_{jk}s_k^A$, where $s_k^A$ is the abundance of the $k$th pollinator. The equations for $d s_k^A/dt$ are obtained by interchanging the labels $P$ and $A$ in the equations and by defining the terms $A^A$, $M^A$ and $W^A$ accordingly.

In this section, we present the exact solution for the structural stability of the models in Eqs.(\ref{eq:fullmf_model})-(\ref{eq:weighted_model}) considering the toy network shown in Fig. 1 in the main text. In order to determine which regions of parameter space are actually accessible by a dynamical system one needs to consider both the stability and the feasibility of the equilibrium solution. The feasibility alone is insufficient as numerical simulations only allow the stable regions to be accessed. The feasibility conditions, that is, positive abundances for all species, should therefore be considered as prerequisite for any equilibrium solution, after which the stability of the equilibrium can be established. 

For the minimal model illustrated in Fig. 1 in the main text, the governing equation for the population dynamics can be written in a matrix form as
\begin{equation}
\frac{ds}{dt}=\left(\alpha^Ts- s^T K s \right),
\label{eq_population_dynalics_matrix}
\end{equation}
We first linearize the system (\ref{eq_population_dynalics_matrix})
around the feasible equilibrium $s^*$. Let us denote a small perturbation around the equilibrium as $ u = s - s^* $. Substituting $ s = u + s^* $ to Eq. (\ref{eq_population_dynalics_matrix}) yields
\begin{equation}\label{key}
\frac{du}{dt} = K^{\text{eff}}u,
\end{equation}
where the entries of matrix $ K^{\text{eff}} $ are given by 
\begin{equation}\label{key}
K^{\text{eff}}_{ij} =\alpha I - K_{ij}s_i^*,
\end{equation}
where $I$ is the identity matrix. By computing the eigenvalues of matrix $ K^{\text{eff}} $ we can retrieve the stability of the 
equilibrium. The stability changes when one of the eigenvalues crosses zero. 
Such points can be found by setting the determinant of matrix $ K^{\text{eff}} $ equal to zero.

\begin{figure}[!htb]
	\centering
	\begin{subfigure}[b]{\textwidth}
		\begin{tikzpicture}[remember picture]
		\node [rectangle, draw, fill=green!20, 
		text width=0.45\textwidth, text centered, rounded corners](identify2) {\\	 
			\textbf{Soft mean-field: Feasibility conditions}
			\begin{multline*}
			s_1^{*},s_2^{*}=\frac{p_1}{q},~~~s_3^{*},s_5^{*}=\frac{p_2}{q},~~~ s_4^{*}=\frac{p_3}{q}\\
			p_1=25 - 2\beta_0^2 +10 \gamma_0 - 3\beta_0 \gamma_0\\
			p_2=25 - \beta_0^2 + 5 \gamma_0 - 2 \beta_0 \gamma_0 - \gamma_0^2\\
			p_3=25 - 5\beta_0 - 2\beta_0^2 + 10\gamma_0 - 2\beta_0\gamma_0+ \gamma_0^2\\
			q=125 +25 \beta_0 - 10\beta_0 ^2 - 2\beta_0^3 -15 \gamma_0^2+4 \beta_0\gamma_0^2
			\end{multline*}
			\textbf{Stability conditions}	
			\begin{equation*}\label{key}
			\gamma_0 = \pm \sqrt{5}\sqrt{5-\beta_0}
			\end{equation*}
			\begin{equation*}\label{key}
			\gamma_0 = \pm \frac{\sqrt{125+25\beta_0-10\beta_0^2-2\beta_0^3}}{\sqrt{15-4\beta_0}}
			\end{equation*}
		};
		\node [rectangle, draw, fill=orange!20, 
		text width=0.45\textwidth, text centered, rounded corners,right of = identify2, node distance=0.5\textwidth] 
		{\\
			\textbf{Weighted: Feasibility conditions}
			\begin{multline*}
			s_1^{*},s_2^{*}=\frac{p_1}{q},~~~s_3^{*},s_5^{*}=\frac{p_2}{q},~~~ s_4^{*}=\frac{p_3}{q}\\
			p_1=\left(5-\beta_0\right)\left(5+\beta_0+2\gamma_0\right)\\
			p_2=25 - \beta_0^2 + 5 \gamma_0 - 2 \beta_0 \gamma_0 - \gamma_0^2\\
			p_3=25 - \beta_0^2 + 10\gamma_0 - \beta_0\gamma_0+ \gamma_0^2\\
			q=\left(5-\beta_0\right)\left(25 +  10\beta_0 - \beta_0^2 -3 \gamma_0^2\right)
			\end{multline*}
			\textbf{Stability conditions}
			\begin{equation*}\label{key}
			\gamma_0 = \pm \sqrt{5}\sqrt{5-\beta_0}
			\end{equation*}
			\begin{equation*}\label{key}
			\gamma_0 = \pm \frac{\beta_0+5}{\sqrt{3}}
			\end{equation*}
		};
		\end{tikzpicture}
	\end{subfigure}
	\caption{Feasibility and stability conditions for soft-mean field and the weighted competition model for the minimal network.}
	\label{fig_feasibility_stability_conditions}
\end{figure}
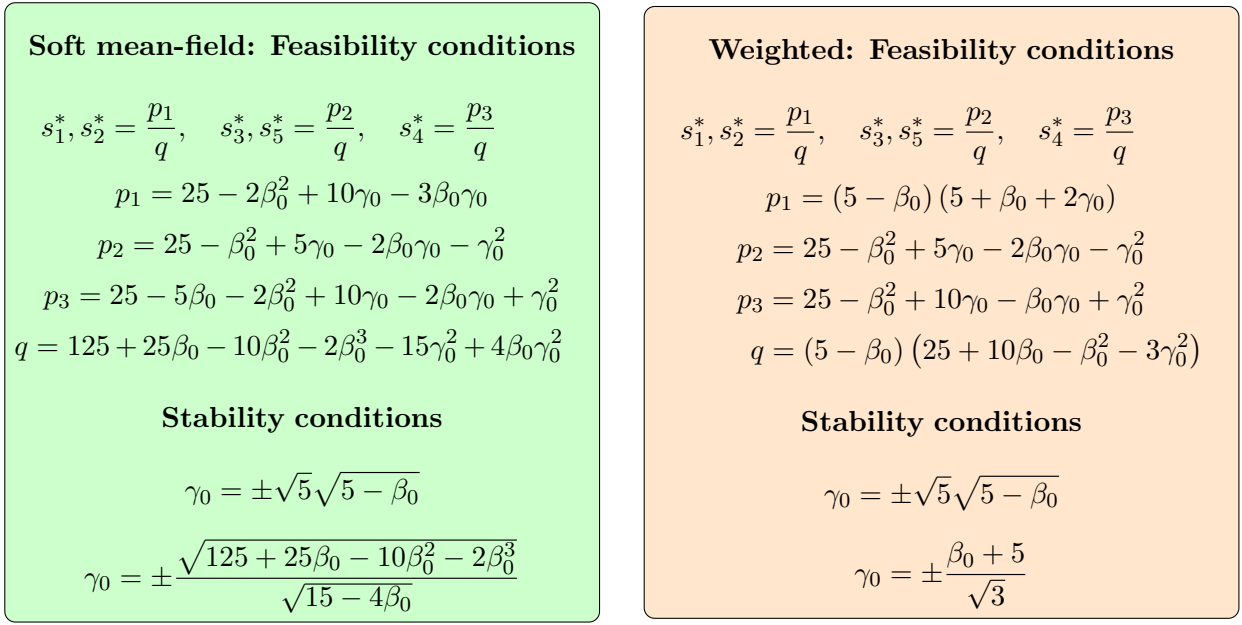

In order to find the stability region of the feasible equilibrium of Eq.~(\ref{eq_population_dynalics_matrix}), with all $s_i>0$, we perform a stability analysis. That is we do not consider the equilibrium with one of  the species having
zero (or negative) abundance. Taking matrix $ K $ of the full mean field as an example, feasible equilibrium is given by the following expressions
\begin{equation}
s_1^{*}=s_2^{*}=\frac{p_1}{q},~~~~s_3^{*}=s_5^{*}=\frac{p_2}{q},~~~~ s_4^{*}=\frac{p_3}{q}
\end{equation}
with
\begin{eqnarray}
p_1=25 +5 \beta_0- 2\beta_0^2 +10 \gamma_0 - 2\beta_0 \gamma_0\nonumber\\
p_2=25 - \beta_0^2 + 5 \gamma_0 - 2 \beta_0 \gamma_0 - \gamma_0^2\nonumber\\
p_3=25 - \beta_0^2 + 10\gamma_0+ \gamma_0^2\nonumber\\
q=125 +50 \beta_0 - 5\beta_0 ^2 - 2\beta_0^3 -15 \gamma_0^2+2 \beta_0\gamma_0^2\nonumber
\end{eqnarray} 
With the feasible equilibrium $ s^* $ and the competition matrix $ K $, we construct the matrix $ K^{\text{eff}} $. Setting the determinant of matrix $ K^{\text{eff}} $ to zero, we find the 
expression for the curves with a vanishing eigenvalue. In the parameter space of $ \beta_0 $ and $ \gamma_0 $, the stability status changes at each time we across those zero lines.
From $ \det  \left(K^{\text{eff}} \right)=0 $, we obtain
\begin{equation*}
\gamma_0 = \pm \left(5-\beta_0\right)
\end{equation*}
\begin{equation*}
\gamma_0 = \pm \frac{\sqrt{125+50\beta_0-5\beta_0^2-2\beta_0^3}}{\sqrt{15-2\beta_0}}
\end{equation*}
which are the boundary curves for stability regions.
Analogously, we describe the exact solution for the 
feasible equilibrium and stability conditions for the
soft mean-field and weighted competition models as shown in Figure~\ref{fig_feasibility_stability_conditions}.

\begin{figure}[h]
	\centering
		\includegraphics[width=\textwidth]{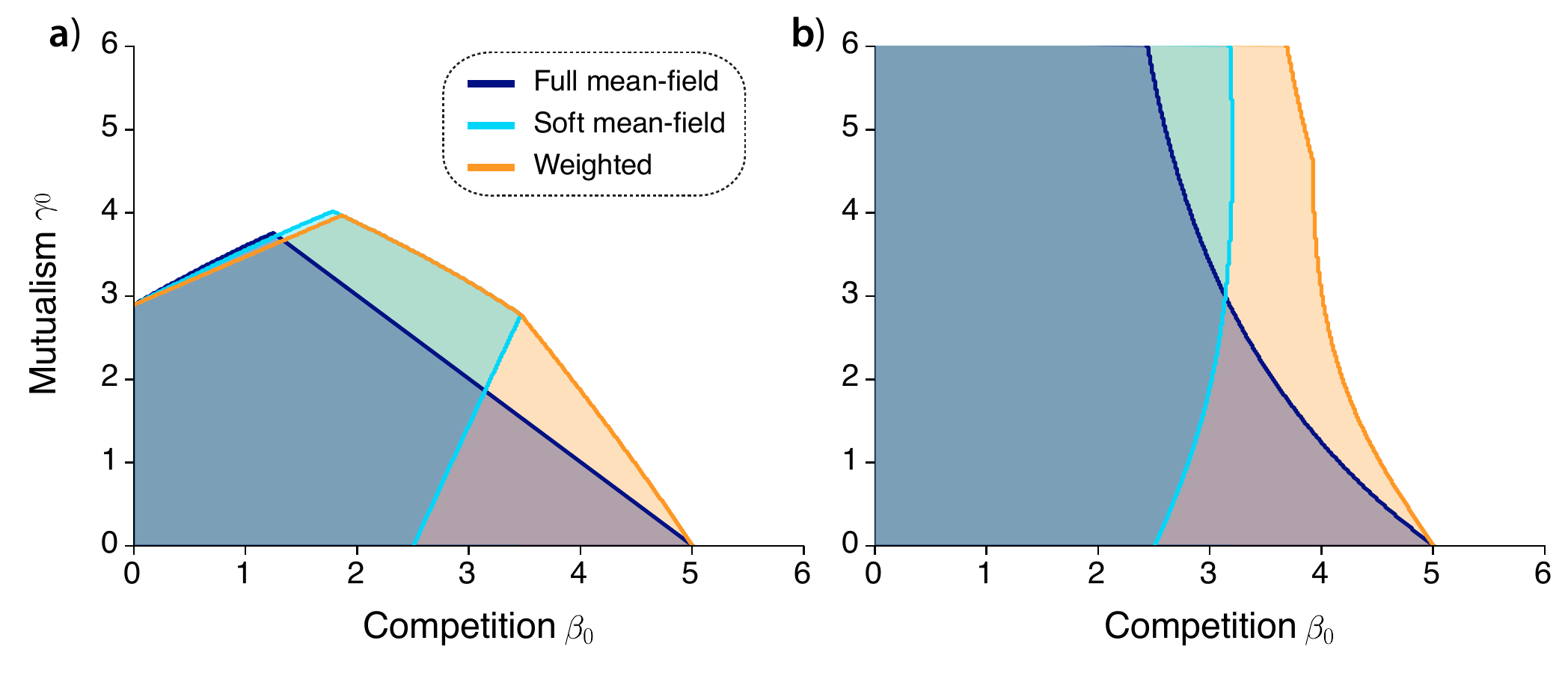}
	\caption{Feasible domain of the minimal model for full mean-field, soft mean-field and the weighted competition. Panel (a) shows the theoretical results for $ h = 0 $, and (b) shows the approximation result for $ h = 0.3 $. Other parameters are taken as $ \alpha_1^P = \alpha_2^A = 1 $, $ \beta^P = \beta^A = 5 $.}
	\label{fig_feasible_domain_toy_model_h03}
\end{figure}

For the case of nonzero Holling term, $h\neq0$, similar feasibility and stability analysis can be established by applying the linear approximation of the mutualistic terms. Figure~\ref{fig_feasible_domain_toy_model_h03} illustrates the analytical results of feasibility and stability conditions for the full mean-field, soft mean-field and weighted competition scenarios, where panel (a) is for $h=0$, and panel (b) is for $h=0.3$. Figure~\ref{fig_feasible_domain_toy_model_h03} suggests an altered feasible area modulated by inter-specific competition for the mutualistic network shown in Figure~\ref{fig_feasibility_stability_conditions}. 

\section{Analytical prediction for the population dynamics of the weighted competition model}
To accurately predict the population dynamics of the weighted competition scenario, we derive the weighted competition matrix. For an observed mutualistic network, and a given initial condition where one realization of population dynamics is sufficient, we derive the competition matrix as 
\begin{equation}\label{eq_competition_matrix_one_realization}
\tilde{A}_{ij}^{P}=\begin{cases}
1 & \textrm{ if }K_{ik}K_{kj}^{T}\neq0\textrm{ for }k\in\arg\min_{k\in A}\left|d_{k}^{A}-\left\lfloor \frac{\sum_{s}K_{is}d_{s}^{A}}{d_{i}^{P}}\right\rfloor \right|,\\
0 & \textrm{otherwise,}
\end{cases}
\end{equation}
where  $d_i^P$ ($d_i^A$) is the number of animals (plants) with which plant (animal) $i$ interacts. Term $ \left\lfloor\frac{\sum_{k}K_{ik}d^A_k}{d^P_i}\right\rfloor$ quantifies the local average of plant competitors mediated by sharing pollinators over number of pollinators $ d^P_i $ of plant $ i $. Recall that in the weighted scenario the competition intensity between two plants (pollinators) is directly proportional to the relative abundance of shared pollinators (plants). Therefore, the more abundant the mutualistic partners, the fiercer the intra-guild competition among the members that have common mutualistic connections. 

\begin{figure}[!tpb]
	\begin{center}
		\includegraphics[width=0.8\linewidth]{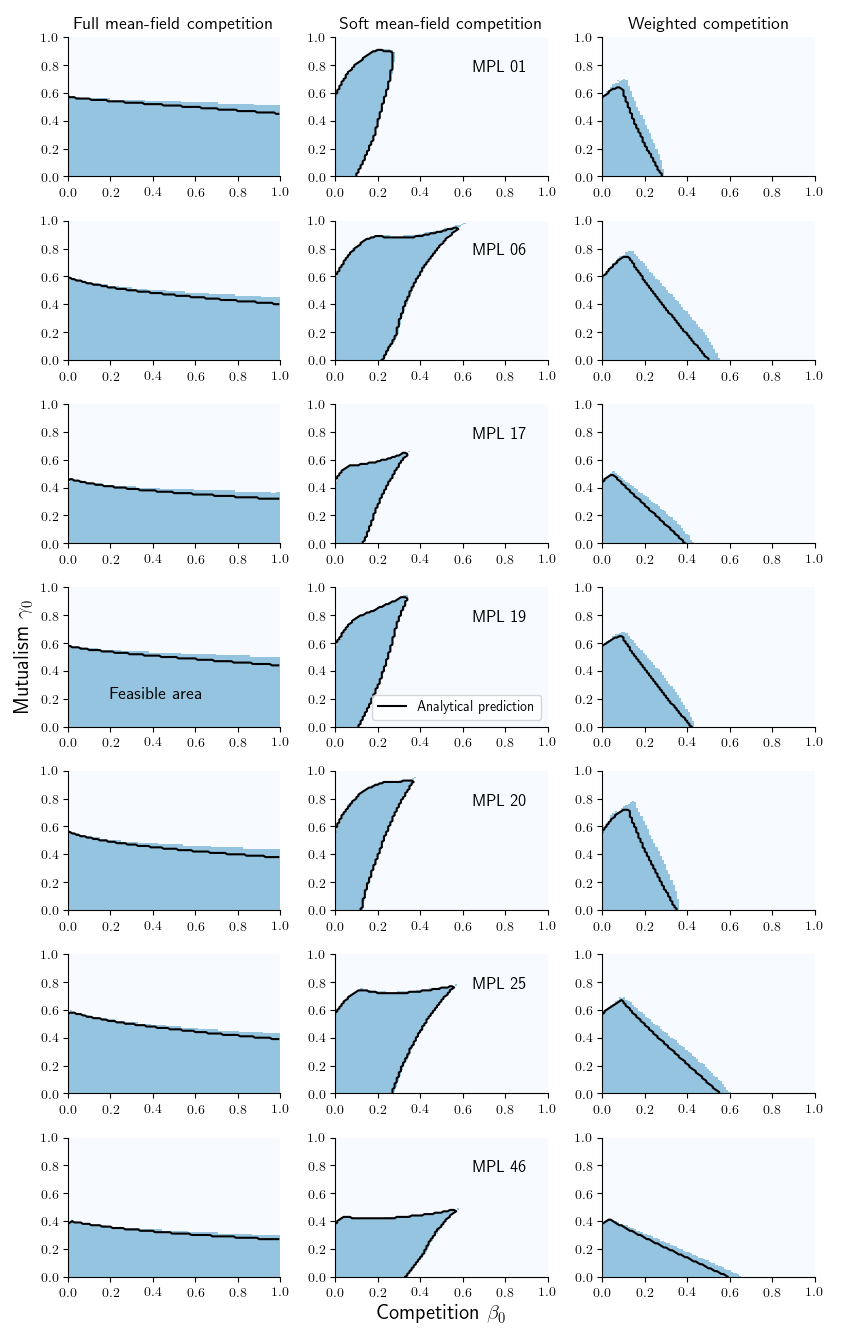}
	\end{center}
	\caption{Feasible area patterns for several networks from the Web of Life platform~\cite{weboflife}, in the (left) full mean-field, (center) soft mean-field and (right) weighted scenarios. Parameters $h=0$, $\beta = 5$, and $\alpha_i^{P,A}=1\;\forall i$. }
	\label{fig:alpha_vs_feasible_area_softMF_h00}
\end{figure}

\begin{figure}[!tpb]
	\begin{center}
		\includegraphics[width=0.8\linewidth]{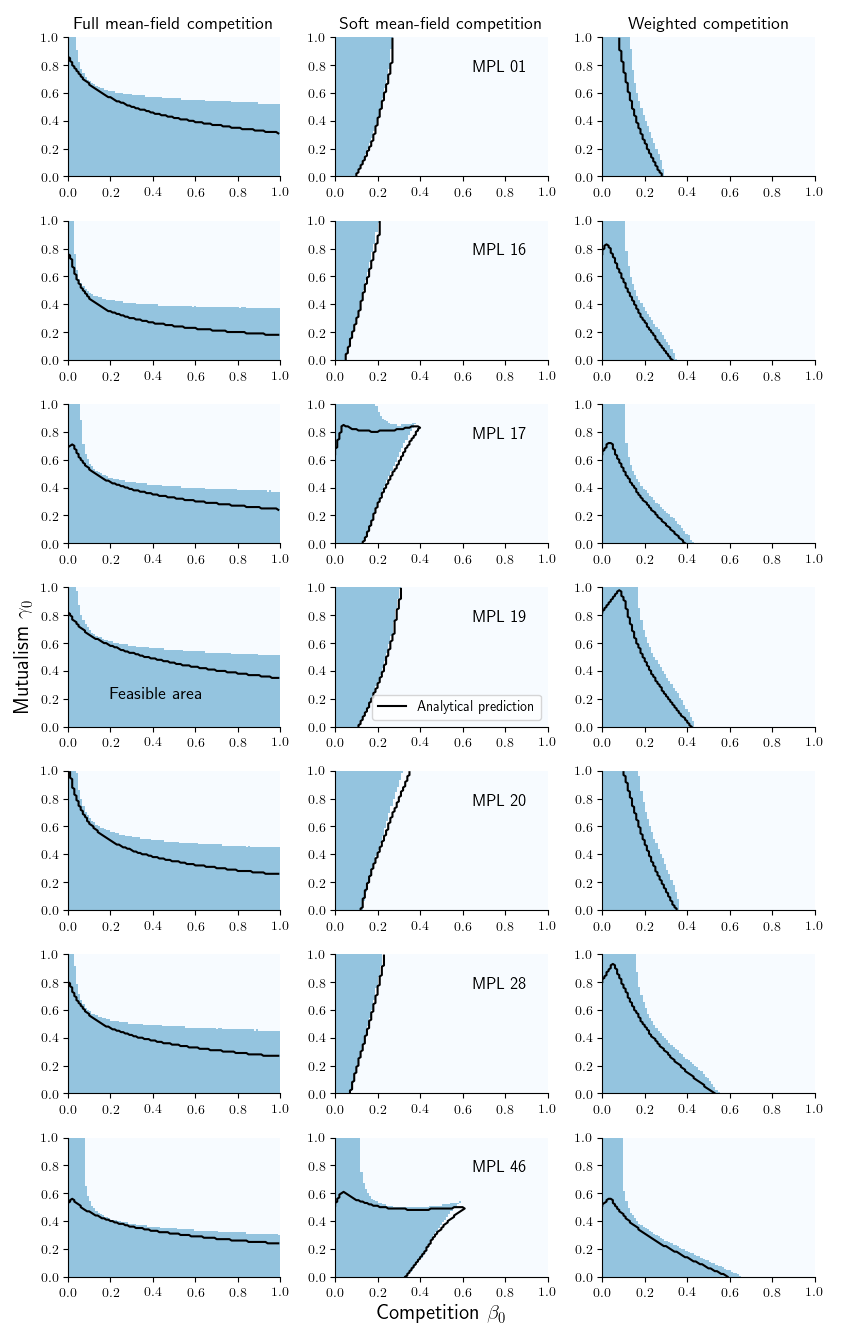}
	\end{center}
	\caption{Feasible area patterns for several networks from the Web of Life platform~\cite{weboflife}, in the (left) full mean-field, (center) soft mean-field and (right) weighted scenarios. Parameters $h=0$, $\beta = 5$, and $\alpha_i^{P,A}=1\;\forall i$.}
	\label{fig:alpha_vs_feasible_area_softMF_h01}
\end{figure}

Figures~\ref{fig:alpha_vs_feasible_area_softMF_h00} and~\ref{fig:alpha_vs_feasible_area_softMF_h01} compare the simulation results of different networks with the analytical predictions. As can be seen, for $h=0$ (Fig.~\ref{fig:alpha_vs_feasible_area_softMF_h00}), the matching between the numerical results and our calculations is remarkable, especially for the full and soft mean-field scenarios. Notice, in particular, how the detailed contour of the feasible area in the latter scenario is almost exactly captured by the analytical curves. As we can see in Fig.~\ref{fig:alpha_vs_feasible_area_softMF_h00}, the feasible area in the weighted competition case is slightly underestimated by the theory, but nonetheless the general trends of the boundaries are reproduced very closely by the solid curves. For the case $h=0.1$ in Fig.~\ref{fig:alpha_vs_feasible_area_softMF_h01} the analytical results accurately predict the feasible area for a large range of mutualistic strength $\gamma_0$, with exceptions at a high value of mutualism due to the saturation effect. All in all, our analytical results work very well for a variety of real plant-pollinator networks. 

\section{Evaluation of analytical predictions on parameterization of population dynamics}

To investigate the performance of our analytical predictions of the feasible area, in this section we present results from extensive simulations considering several parameter combinations. Specifically, we compare the analytical solutions derived in the previous section and in the main text
for real plant-pollinator networks under different choices for the intraspecific competition $ \beta $ and the Holling term $ h $.

\subsection{Variation in the intraspecific competition $ \beta$}
\begin{figure}[h]
	\centering
		\includegraphics[width=\textwidth]{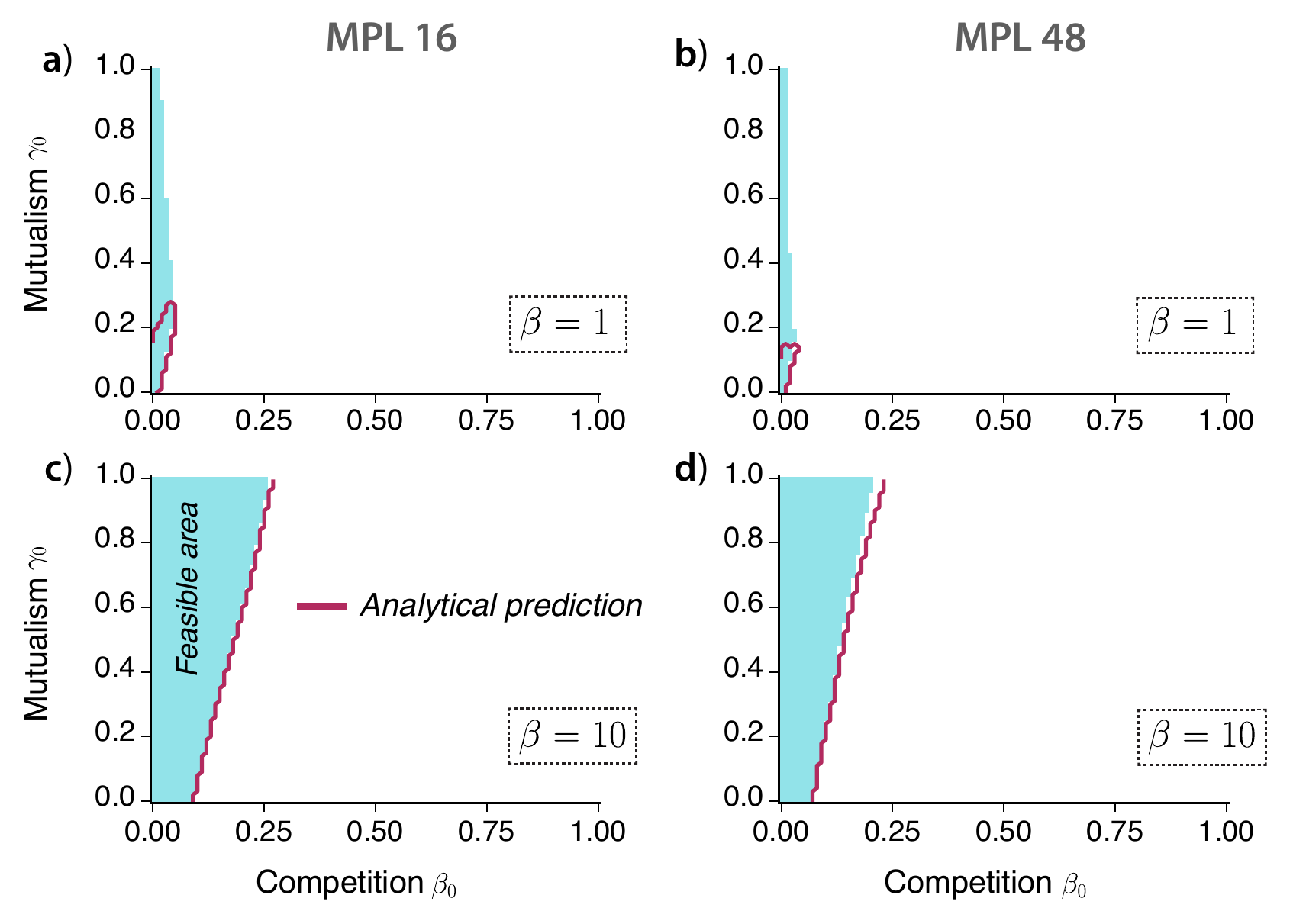}
	\caption{Analytical prediction (black curve) for the soft mean-field, considering different values for the inter-specific competition strength $\beta_i^A = \beta_i^P = \beta$. The system in Eq.~(\ref{eq:softmf_model}) was numerically integrated with the Heun's method, considering total simulation time $T=2000$ and time step $dt = 0.01$. The simulation result (shaded area) is obtained with parameters $ \alpha_i^P = \alpha_i^A = 1 $ and $ h = 0.1 $. }
	\label{fig_feasible_beta_gamma_approx_meanField}
\end{figure}

\begin{figure}[h]
	\centering
		\includegraphics[width=\textwidth]{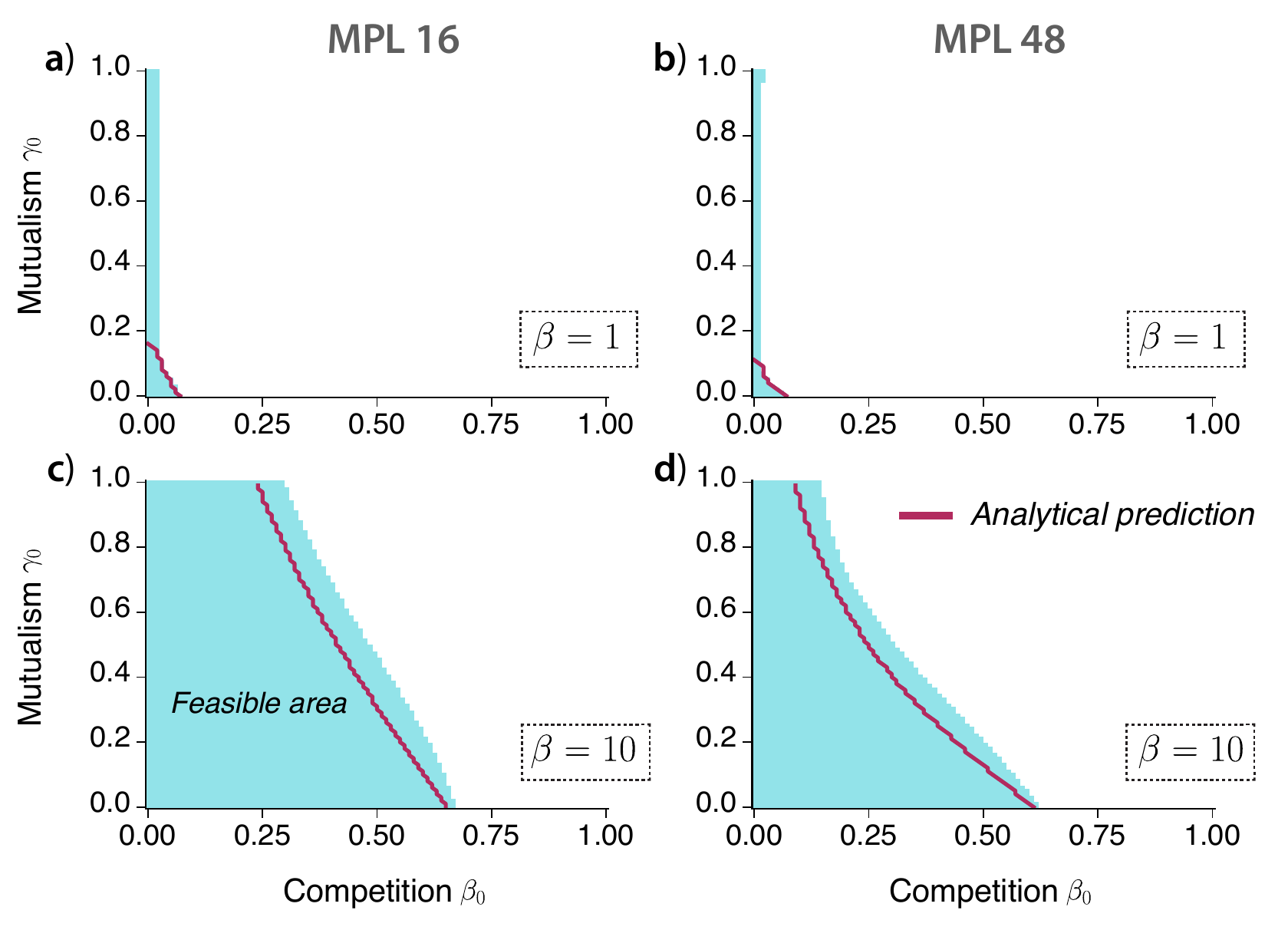}
	\caption{Analytical prediction (black curve) for the weighted competition scenario, considering different values for the inter-specific competition strength $\beta_i^A = \beta_i^P = \beta$. The systems in Eqs.~(\ref{eq:weighted_model}) was numerically integrated with the Heun's method, considering total simulation time $T=2000$ and time step $dt = 0.01$. The simulation result (shaded area) is obtained with parameters $ \alpha_i^P = \alpha_i^A = 1 $ and $ h = 0.1 $.}
	\label{fig_feasible_beta_gamma_approx_weighted}
\end{figure}

Figures \ref{fig_feasible_beta_gamma_approx_meanField} and \ref{fig_feasible_beta_gamma_approx_weighted} show the numerical and analytical results when we increase or decrease the intensity of intraspecific competition $ \beta_i $ for the observed mutualistic networks \emph{M-PL-048} and \emph{M-PL-016}. For the population dynamics with soft mean-field competition, the analytical results accurately predict the feasible area for both the increase and decrease of intraspecific competition $ \beta $. For the weighted competition scenario, the linear approximation captures well the changing behavior of species coexistence. 
\subsection{Weak mutualism $ h \gg 1 $}
Compared to the majority of the theories in ecology dealing with either weak mutualism ($ h \gg 1 $) or strong mutualism $ h =0 $, our theoretical framework fills the gap and covers the full range ($ 0 < h < 1$). We proceed by presenting the feasible area in the extreme cases of $ h = 0 $ and $ h \gg 1 $. 

For a strong mutualism $ h =0 $, the dynamical model is reduced to Lotka--Volterra equations with type II functional responses. Strong mutualism without saturation is often unstable leading to indefinite and unbounded growth of species which is argued to be biologically unrealistic due to the environmental constraints like carrying capacity \cite{wright1989simple}. However, strong mutualism at the same time overpowers the inter-specific competition resulting in a cut-off of the mutualism intensity $ \gamma_0 $.

\begin{figure}[h]
	\centering
		\includegraphics[width=\textwidth]{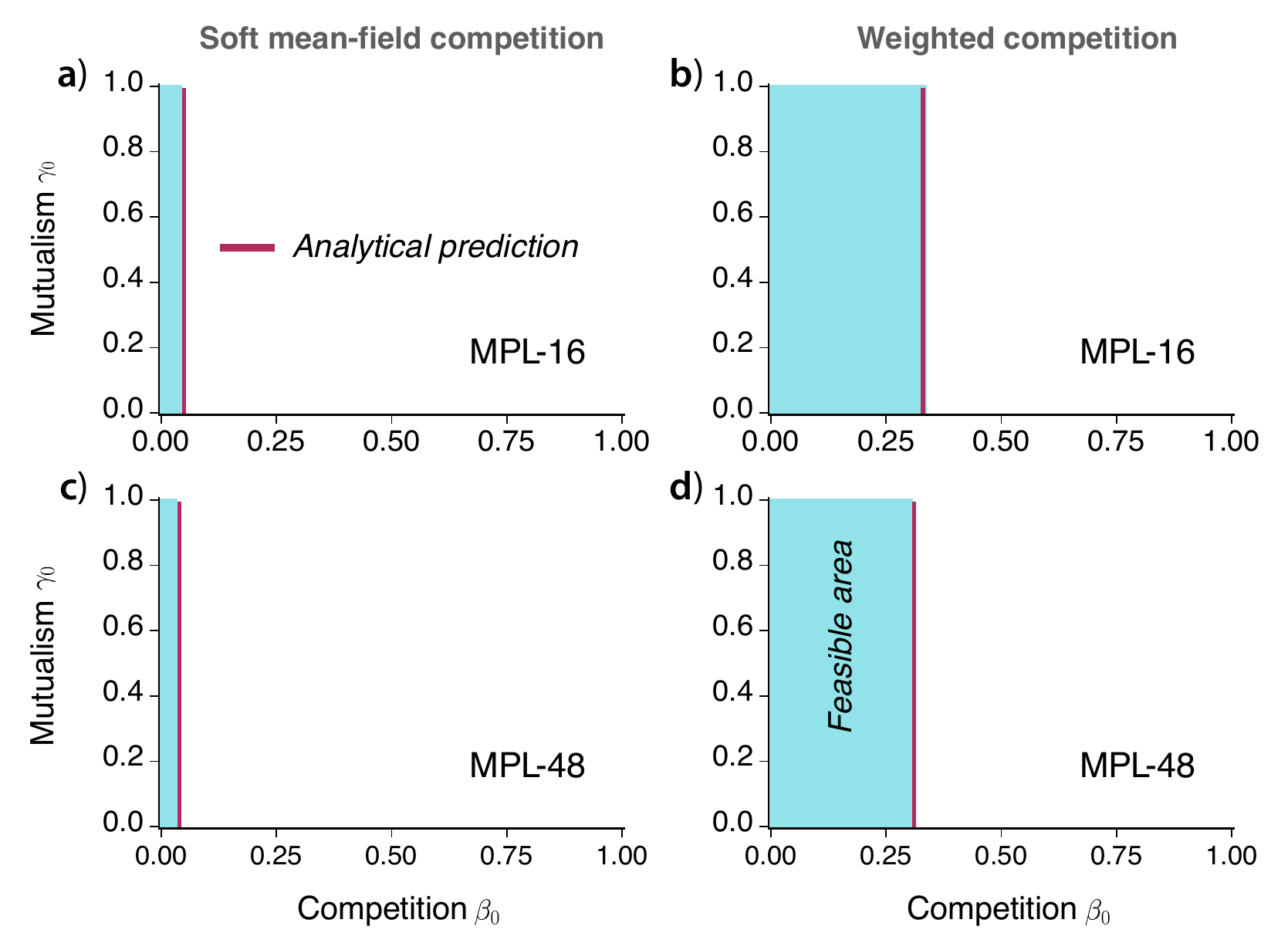}
	\caption{Performance of analytical predictions (black lines) in the weak mutualism regime ($ h \gg 1  $). Other parameters: $\alpha_i^P = \alpha_i^A = 1$, $\beta = 5$, and $h=10^3$. The systems in Eqs.~(\ref{eq:softmf_model}) and~(\ref{eq:weighted_model}) were numerically integrated with the Heun's method, considering total simulation time $T=2000$ and time step $dt = 0.01$.}
	\label{fig_feasible_beta_gamma_approx_multilayerh1000}
\end{figure}

For a weak mutualism $ h \gg 1 $, the mutualism term is saturated to a constant $ 1/h $, i.e., 
\begin{equation}\label{key}
\lim\limits_{ h \to \infty}\frac{\gamma_0 M_i^P}{1+ h\gamma_0 M_i^P} \approx \frac{1}{h}
\end{equation}
When the mutualism is saturated, the dynamical model is reduced to a linear model, and the feasible equilibrium is obtained by solving
\begin{equation}\label{eq_feasibbe_beta_gamma_asymptotic}
\begin{bmatrix}
s^P \\
s^A
\end{bmatrix} =\left(\beta I + \beta_0
\begin{bmatrix}
A^P &0\\
0& A^A
\end{bmatrix}
\right)^{-1}
\begin{bmatrix}
\alpha^P+\frac{1}{h}u\\
\alpha^A+\frac{1}{h}u
\end{bmatrix}
\end{equation}
where $ u $ is the all-one vector. For the weighted competition case, a linear approximation on the matrices $ \tilde{A}^P $ and $ \tilde{A}^A $ is applied. The mutualism intensity $ \gamma_0 $ is disentangled from affecting the feasible area (which is confirmed by the numerical results in Fig.~\ref{fig_feasible_beta_gamma_approx_multilayerh1000}). Moreover, saturated mutualism, in turn, imposes limitations on the intensity $ \beta_0 $ of competition. Because without the compensation of mutualism benefit, the intra-specific competition might be overly strong and eventually make species go extinct.   

\section{Structural stability and mutualistic network properties}
In this section, we provide complementary results on the correlation between feasible area and network architectures for dynamical models of full mean field, soft mean field and weighted competition. We test two more network properties, namely, the variance of the degree sequence and the ratio of inter-degree to intra-degree for both plants and animals. In addition, we test the maximum interspecific competition $ \beta_0 $ before losing any species and its relation to network architectures when the mutualism is saturated.

\begin{figure}[!h]
	\centering
		\includegraphics[width=\textwidth]{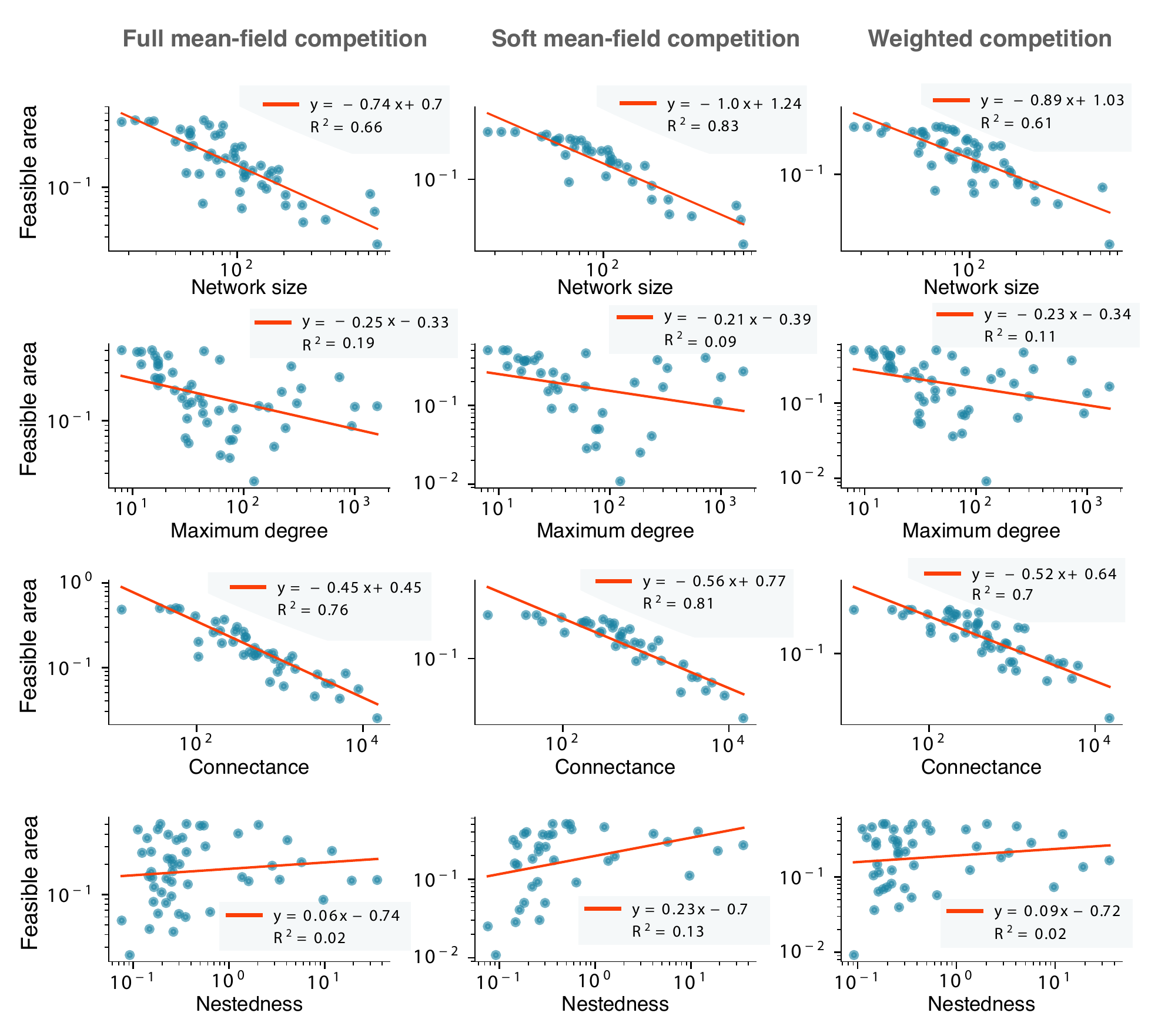}
	\caption{Feasible area and network architecture. Simulations are performed on $ 50 $ real-world mutualistic networks. Parameters to compute the feasible area are $ \alpha = 1 $, $ \beta = 5 $, $ h = 0 $, $ \beta_0 \in [0,1] $ and $ \gamma_0  \in [0,1]$.}
	\label{fig_netproperty_feasible_areah0}
\end{figure}

\begin{figure}[!h]
	\centering
		\includegraphics[width=\textwidth]{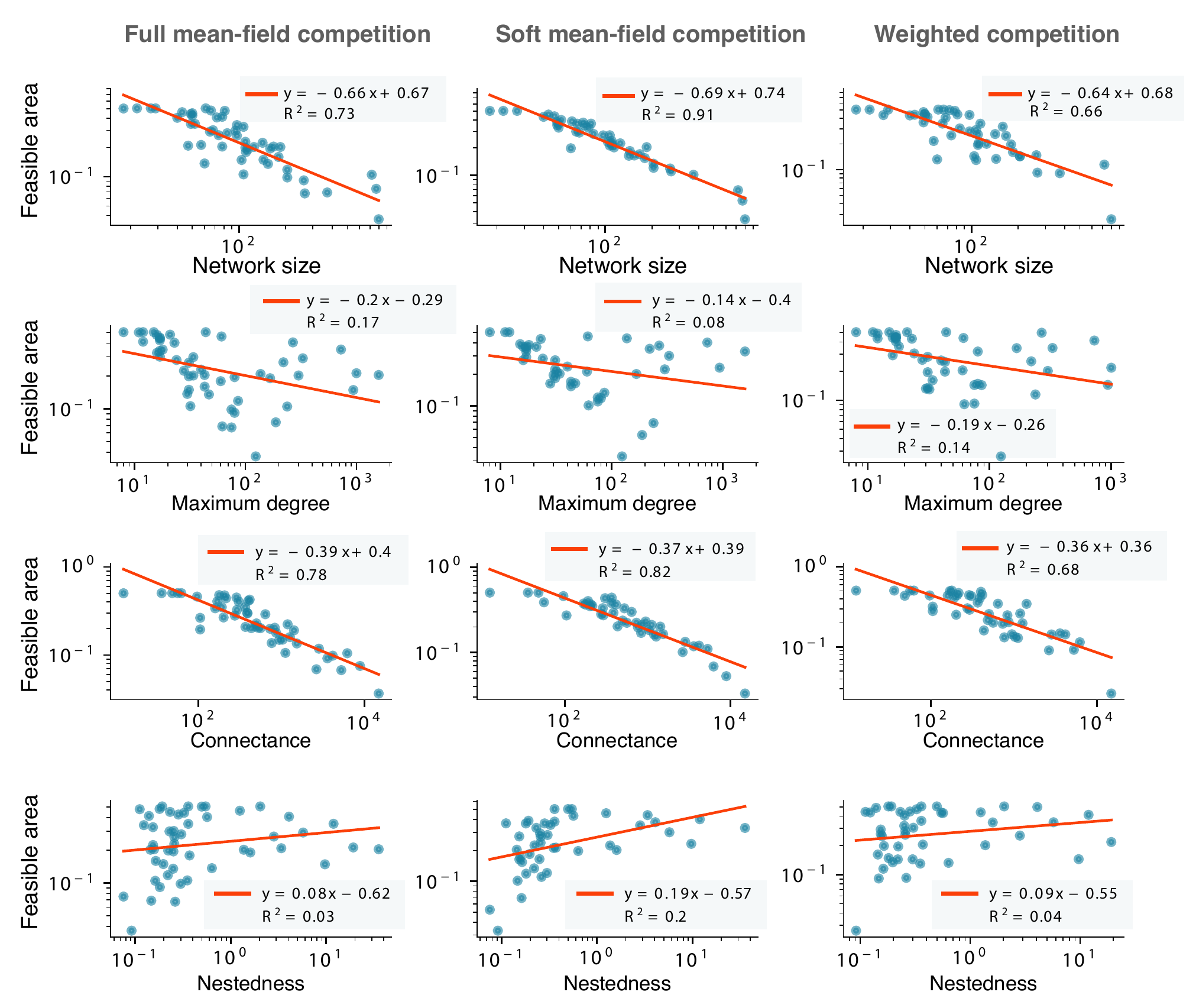}
	\caption{Feasible area and network architecture. Simulations are performed on $ 50 $ real-world mutualistic networks. Parameters to compute the feasible area are $ \alpha = 1 $, $ \beta = 5 $, $ h = 0.1 $, $ \beta_0 \in [0,1] $ and $ \gamma_0  \in [0,1]$.}
	\label{fig_netproperty_feasible_areah01}
\end{figure}

Here, we provide the correlation between feasible area and network architectures for all the considered dynamical models. In particular, we consider the regimes of strong ($ h = 0 $) and saturated mutualism ($ h \neq 0 $). For the strong mutualism regime (Fig.~\ref{fig_netproperty_feasible_areah0}), global network properties, such as the number of species and connectance, correlate strongly and negatively with the feasible area, with goodness-of-fit of $ R^2 = 0.83  $ and $ R^2 = 0.81 $ for the soft mean-field competition. The maximum degree shows a less strong correlation with the feasible area, and neither does nestedness show a clear dependence on the same quantity. The same trends are observed
for other competition scenarios in the strong mutualism regime (see Fig.~\ref{fig_netproperty_feasible_areah0}) as well as for the dynamics with saturated mutualism (Fig.~\ref{fig_netproperty_feasible_areah01}). 

\section{Analysis on the diversity-stability relation}

May~\cite{may1972will} has established a stability-diversity relation for a large ecological system whose stability is characterized by $\frac{dx}{dt} = J x$, indicating that a large system is less stable. In May's assumption, each element of the Jacobian matrix $ J $ is equally likely to be positive or negative, having an absolute magnitude chosen from a random distribution with zero mean and standard deviation $ \alpha $. Matrix $ J $ can also be written in the form $ J = B-I $. Based on Theorem \ref{theorem_rand_mat_eig} for the largest eigenvalue of a random matrix \cite{furedi1981eigenvalues}, the system is stable if
\begin{equation}\label{may_condition}
\sigma\sqrt{NC}-1 \leq 0
\end{equation}
where $\sigma $ is the standard deviation of the random variable from which the off-diagonal elements of the Jacobian matrix $ J $ take value; $ N $ is the size of the network, and $ C $ denotes the network connectance~\cite{furedi1981eigenvalues}. 

\begin{theorem}\label{theorem_rand_mat_eig}
	Let $ M $ be a random $ N \times N $ real and symmetric matrix where elements $ m_{ij} = m_{ji}  $ are independent random variables. Assume that these random variables possess a common mean $ E[m_{ij}]=0 $ and common variance $ \text{Var}[m_{ij}] = \sigma^2 $ and $ E[m_{ii}]=\mu $. Then the largest eigenvalue is upper bounded by 
	\begin{equation}\label{key}
	\max |\lambda_1| \leq 2 \sigma \sqrt{N} +O(N^{1/3}\log N)
	\end{equation}
	where $ \sigma $ is the standard variation.
\end{theorem}

Our goal now is to examine how the stability condition in Eq.~\ref{may_condition} relates with the feasible 
area of real plant-pollinator networks. However, 
as discussed in the main text, Eq.~(\ref{may_condition}) is
associated with the probability that the system is stable given 
a particular set of parameters; the feasible area, on the other hand, is defined for a set of parameters. Therefore, in order to establish a relation between May's condition and the feasible area, we define  
\begin{equation}
C_{\mbox{\small{May}}}=\langle\langle J_{ii}\rangle\rangle_{(\beta_0,\gamma_0)}-\left\langle \sigma(J_{ij})\right\rangle _{(\beta_0,\gamma_0)}\sqrt{NC},
\label{eq:conditions_betagamma}
\end{equation}
where $\langle \cdot \rangle$ is an average over the Jacobian matrix's elements,
and  $\langle  \cdot  \rangle_{(\beta_0,\gamma_0)}$ corresponds to the average over certain ranges of parameters $\beta_0$ and $\gamma_0$. The first term in Eq.~(\ref{eq:conditions_betagamma}), $\langle\langle J_{ii} \rangle \rangle_{(\beta_0,\gamma_0)}$, is the average taken over the diagonal elements, since, contrarily to the random model considered by May, the diagonal elements of the Jacobian, $J_{ii}^{\rm{P,A}}$, are not constant, but rather are heterogeneously distributed over the diagonal (see Appendix); the term $\langle \sigma(J_{ij})  \rangle_{(\beta_0,\gamma_0)}$ is the average standard deviation of the off-diagonal values of $\mathbf{J}$.

\begin{figure}[!htp]%
	\centering
		\includegraphics[width=\textwidth]{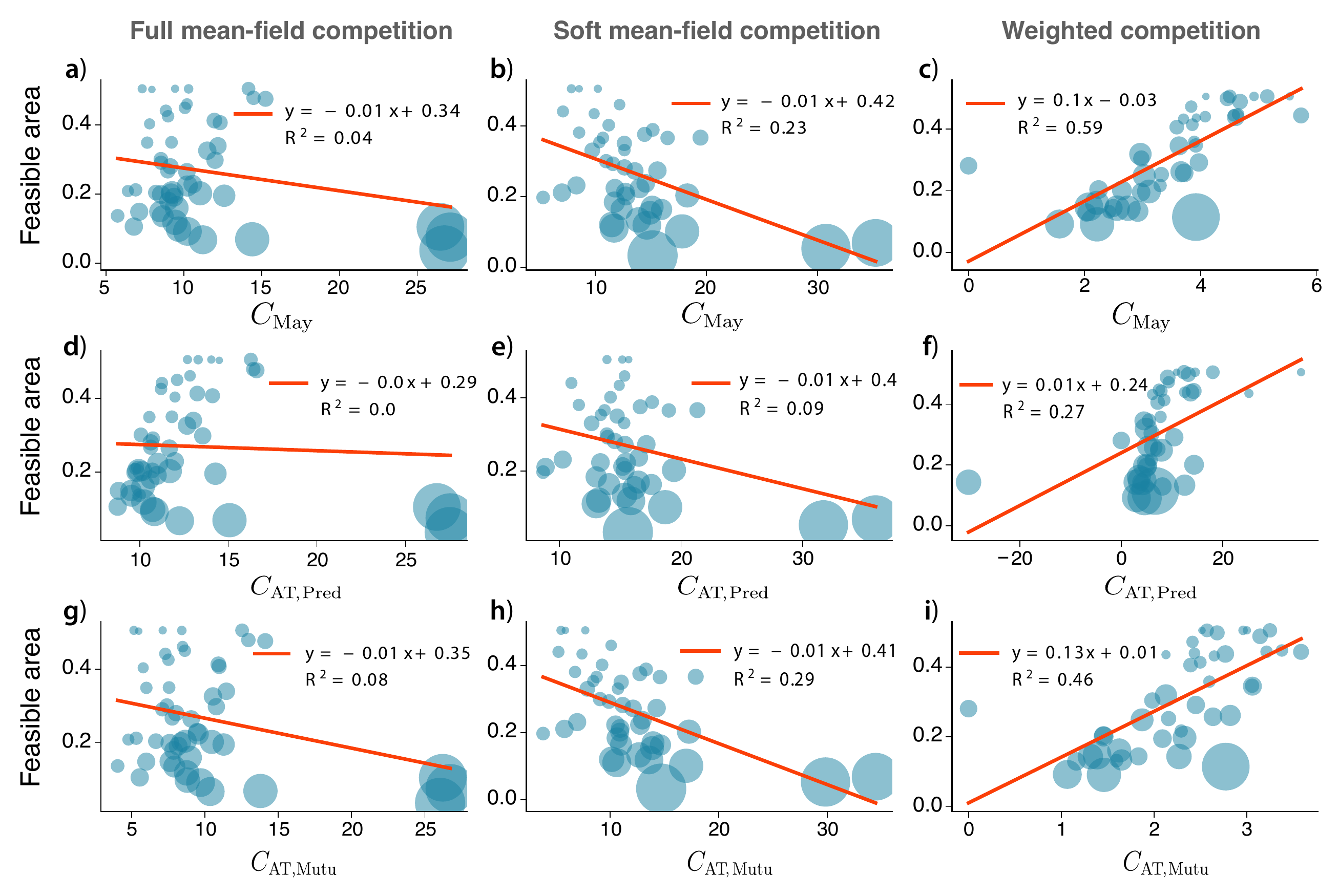}%
	\caption{Correlation of feasible area with (a)-(c) May's stability  criterion [Eq.~(\ref{eq:conditions_betagamma})], (d)-(f) stability criterion by Allesina and Tang~\cite{allesina2012stability} for Predator-Prey  models [Eq.~(\ref{eq:alletang_pred})], and (g)-(i) criterion by Allesina and Tang for random matrices with competitive and mutualistic interactions [Eq.~(\ref{eq:alletang_mutu})]. Dots correspond to the real plant-pollinator networks with indexes between 01 and 50 retrieved from the Web of Life database~\cite{weboflife}. Size of the dots is proportional to the network size. The dynamics of all networks was numerically integrated with the Heun's method, considering total simulation time $T=1000$ and a time step $dt=0.01$. The feasible area was calculated over a $\beta_0 \times \gamma_0 $ grid with $100 \times 100$ points, where $\beta_0,\gamma_0 \in [0,1]$. Other parameters: $h=0.1$, $\alpha_i=1$ $\forall i$, $\beta=5$. Solid lines correspond to the linear least-square regression, and $R^2$ is the correlation coefficient.}
	\label{fig_correlation_farea_May}%
\end{figure}

\begin{figure}
	\centering
		\includegraphics[width=\textwidth]{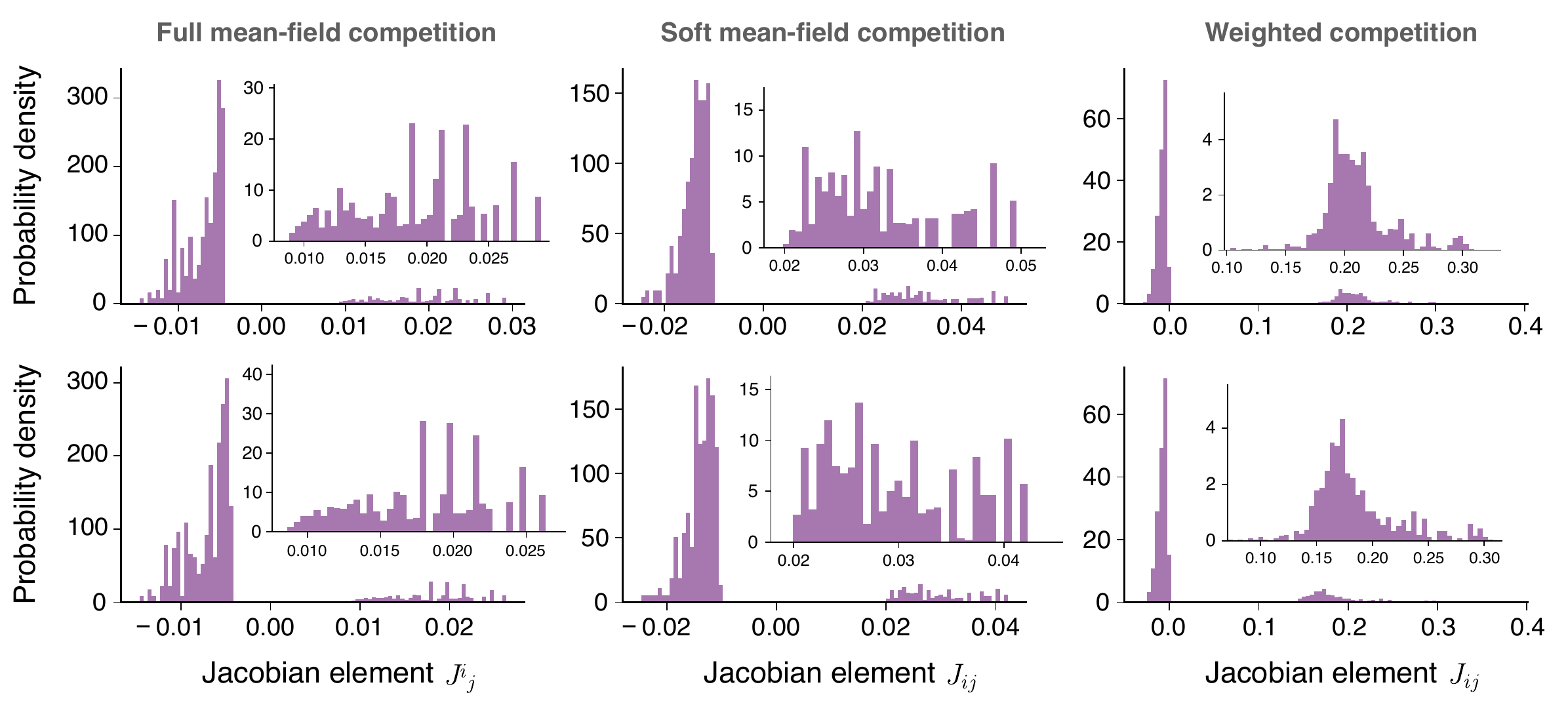}%
	\caption{Distribution of the elements of the Jacobian matrix for network MPL-10 for (upper row) $h=0$ and (lower row) $h=0.1$. Parameters $\alpha =1$, $\beta=5$, $\beta_0 = 0.1$, $\gamma_0=0.2$.}
	\label{fig_distri_Jacobian_MPL010}
\end{figure}

\begin{figure}[!h]
	\centering
		\includegraphics[width=\textwidth]{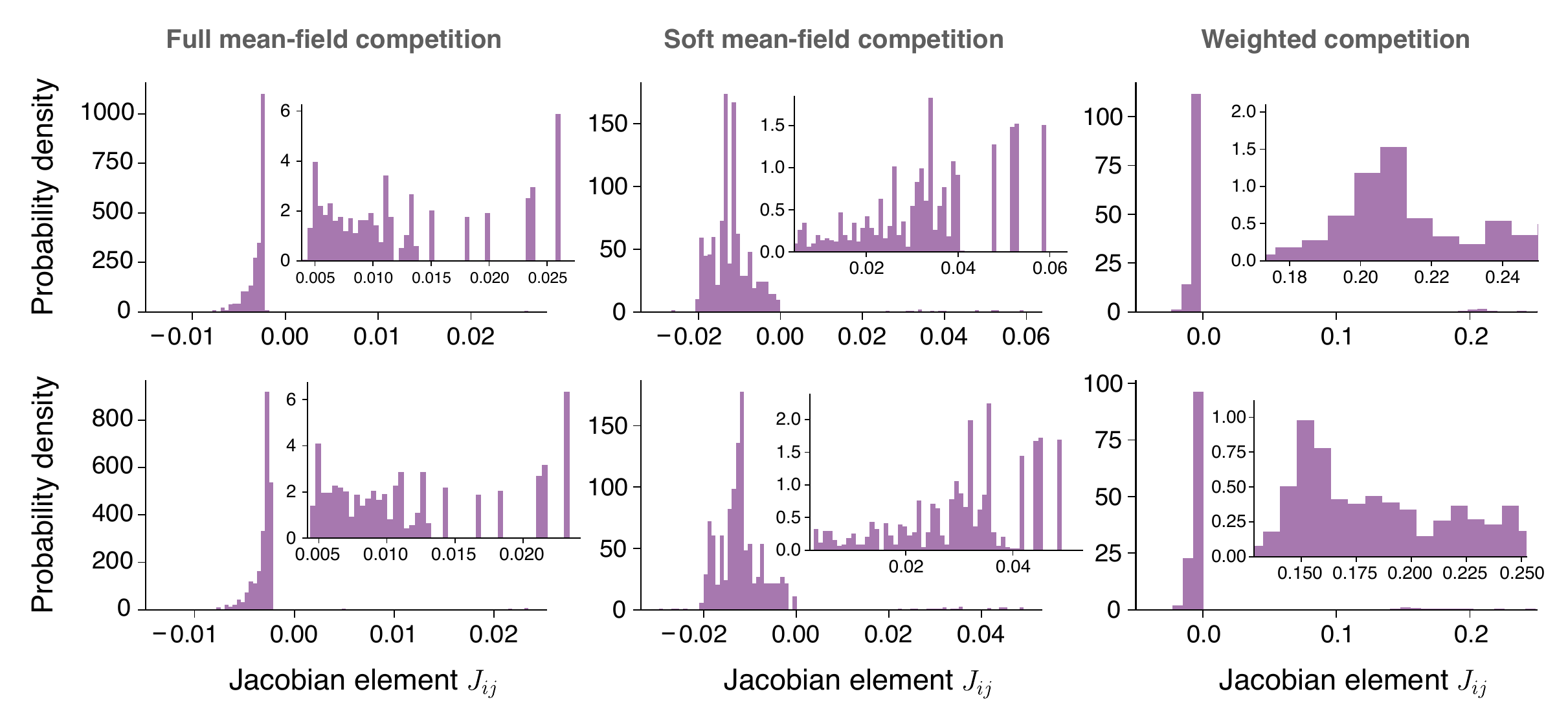}
	\caption{Distribution of the elements of the Jacobian matrix for network MPL-48 for (upper row) $h=0$ and (lower row) $h=0.1$. Parameters $\alpha =1$, $\beta=5$, $\beta_0 = 0.1$, $\gamma_0=0.2$.}
	\label{fig_distri_Jacobian_MPL048}
\end{figure}

The stability criterion by May is derived considering purely random interactions among the elements in a complex system, i.e., the Jacobian matrix is a random matrix without any constraints on its elements. More recently, Allesina and Tang (AT)~\cite{allesina2012stability} generalized May's stability analysis for random matrix models that incorporate aspects of predator-prey, mutualisc and competitive interactions~\cite{allesina2012stability}. Seeking to verify how these more complex stability criteria relate with the feasible area of networks, we define
\begin{align}
C_{\mbox{\small{AT,Pred}}}&=\langle\langle J_{ii}\rangle\rangle_{(\beta,\gamma)}-\left\langle \sigma(J_{ij})\right\rangle _{(\beta,\gamma)}\sqrt{NC}\left(1-\frac{\left\langle \mathbb{\mathbb{E}}^{2}\left(|J_{ij}|\right)\right\rangle _{(\beta,\gamma)}}{\left\langle \sigma(J_{ij})\right\rangle _{(\beta,\gamma)}^{2}}\right)\label{eq:alletang_pred}\\
C_{\mbox{\small{AT,Mutu}}}&=\langle\langle J_{ii}\rangle\rangle_{(\beta,\gamma)}-\left\langle \sigma(J_{ij})\right\rangle _{(\beta,\gamma)}\sqrt{NC}\left(1+\frac{\left\langle \mathbb{\mathbb{E}}^{2}\left(|J_{ij}|\right)\right\rangle _{(\beta,\gamma)}}{\left\langle \sigma(J_{ij})\right\rangle _{(\beta,\gamma)}^{2}}\right)\label{eq:alletang_mutu},
\end{align}
where $C_{\mbox{\small{AT,Pred}}}$ corresponds to the stability condition of random predator-prey matrices~\cite{allesina2012stability}, and $C_{\mbox{\small{AT,Mutu}}}$ is the analogous quantity for random matrix models that emulate the interplay between competitive and mutualistic relationships~\cite{allesina2012stability}. Figure~\ref{fig_correlation_farea_May} shows the observed correlation of May's and Allesina and Tang's stability with the feasible area for full mean-field, soft mean-field and weighted competition scenarios. Interestingly, we do not observe a significant dependence of the feasible area on conditions (\ref{eq:conditions_betagamma})-(\ref{eq:alletang_mutu}) for full and soft mean-field scenario. A pattern does emerge in the weighted scenario, for which we have a positive correlation between feasible area and all conditions; however, it is interesting to note that the strongest correlation occurs for $C_{\mbox{\textrm{\small{May}}}}$. The surprise about this result resides in the fact that $C_{\mbox{\textrm{\small{May}}}}$ is the condition derived for the simplest random matrix model. At first sight, one would expect to observe a more significant dependence of the feasible area on $C_{\mbox{\small{AT,Mutu}}}$, which accounts for competitive and mutualistic networks, but what we have is the opposite: the most complex model (weighted competition scenario) adheres best with the condition derived from May's stability criteria. As we argue in the main text, the explanation for the latter results lies in the expression of the Jacobian elements (see Appendix). In the weighted competition scenario, the off-diagonal Jacobian elements $J_{ij}$ depend on terms $1/s_{i}^{P,A}$ and $1/(s_{i}^{P,A})^2$. Since the abundance values are generally less than 1, the elements $J_{ij}$ end up being narrowly peaked around an average value, thus making the standard deviation $\sigma(J_{ij})$ to be less significant than the average
off-diagonal terms $J_{ii}$. To exemplify this phenomenon, in Figs. \ref{fig_distri_Jacobian_MPL010} and \ref{fig_distri_Jacobian_MPL048} we show the distribution of elements in Jacobian matrix for networks MPL-010 and MPL-048, respectively.

In order to get further insights into the dynamics of the three competition models, let us denote the right hand side of Eq. (\ref{eq:weighted_model}) as $ f(s_i^P) $.
To determine the stability of the equilibrium point, we analyze the Jacobian matrix $ J $ which can be written as
\begin{equation}\label{key}
J = B - \beta I 
\end{equation}
where $ B $ can be expressed in the form of block matrix as
\begin{equation}\label{key}
B = \begin{bmatrix}
\left(B_{11} \right)_{N^P \times N^P}& \left(B_{12} \right)_{N^P \times N^A} \\
\left(B_{21} \right)_{N^A \times N^P} & \left(B_{22} \right)_{N^A \times N^A} \\
\end{bmatrix}
\end{equation}
Each element in $B_{11}$ is calculated by $ \left( B_{11}
\right)_{iu}  = \frac{\partial f(\left(s^*\right)_i^P)}{\partial 
\left(s^*\right)_u^P} $, where $ s^* $ is the abundance at 
equilibrium. Each element in $B_{12}$ is calculated by $ \left( 
B_{12}\right)_{iv}  = \frac{\partial f(\left(s^*\right)_i^P)}
{\partial \left(s^*\right)_v^A} $. An analogous form can be 
obtained for $ B_{22} $ and $ B_{21} $ by replacing superscript 
P indicating plant species to superscript A representing 
pollinator species and vice versa. We derive the expression of 
submatrices $ B_{11} $ and $ B_{12} $ for three cases of full 
mean-filed, soft mean-field and weighted competition (see also Appendix~\ref{sec:appendix}).

(i) Case of full mean-field competition: the submatrix $ B_{11} = \beta_0J_{N^P}$, where $ J_{N^P} $ is the all one matrix. Therefore, all elements in Jacobian submatrix $ B_{11} $ are linearly correlated. Each element in $ B_{12} $ is determined mainly by mutualistic interactions $ \left( B_{12}\right)_{iv} = \frac{\gamma_0K_{iv}}{\left(1+h \gamma_0 \sum_k K_{ik}\left(s^*\right)_k^A\right)^2} $. All the interacting pollinators of plant $ i $ have the same value in Jacobian submatrix $ B_{12} $ and thus linearly correlated.

(ii) Case of soft mean-field competition: the submatrix $ B_{11} $ has an element of $ 1 $ whenever there is competition which has the same value for all competed species. Each element in $ B_{12} $ is determined mainly by mutualistic interactions $ \left( B_{12}\right)_{iv} = \frac{\gamma_0K_{iv}}{\left(1+h \gamma_0 \sum_k K_{ik}\left(s^*\right)_k^A\right)^2} $.

(iii) Case of weighted competition: the submatrix $ B_{11} $ has elements computed by
\begin{equation}\label{key}
\left(B_{11}\right)_{iu}  = \beta_0 \frac{\sum_k K_{ik}K^T_{ku}\left(s^*\right)_k^A }{\sum_k K_{ik}\left(s^*\right)_k^A } 
\end{equation}
For each pollinator $ u $ of plant $ i $, the value in Jacobian submatrix is varied, in contrast to the same value of $ 1 $ in the case of full mean field and soft mean field. Each element in submatrix $ B_{12} $ is computed by 
\begin{equation}\label{eq_B12_multilayer}
\left(B_{12}\right)_{iv} =  \frac{\beta_0 \sum_{j}K_{iv}K_{vj}\left(s^*\right)_j^P\sum_{k\neq v} K_{ik}\left(s^*\right)_k^A  }{\left(\sum_k K_{ik}\left(s^*\right)_k^A \right)^2}+\frac{\gamma_0K_{iv}}{\left(1+h \gamma_0 \sum_k K_{ik}\left(s^*\right)_k^A\right)^2} 
\end{equation}
The second term in the right hand side of Eq. (\ref{eq_B12_multilayer}) is introduced due to mutualistic interactions and shows an analogous pattern to the full mean field case and soft mean field case. However, the first term uniquely appears in the weighted competition scenario. In addition, the value is varied for different pollinators $ v $, determined by the number of introduced plant competitions $ \sum_{j}K_{iv}K_{vj}\left(s^*\right)_j^P $ mediated by sharing a common pollinator $ v $. Weighted competition reduces the correlation between elements in each row of the Jacobian submatrix $ B_{12} $ and, therefore, shows a well agreement with May's stability criteria, which is built upon the assumption of independence among elements of the Jacobian matrix. In the Appendix A we provide the complete expressions for the elements of the Jacobian matrix.


\appendix

\section{Expressions for the Jacobian matrix elements}
\label{sec:appendix}
The Jacobian matrices of the models considered in the main text can be expressed as follows
\begin{equation}
J=\left(\begin{array}{cc}
J^{P} & J^{PA}\\
J^{AP} & J^{A}
\end{array}\right).
\end{equation}
In the sequel we write the expression for each competition scenario.

\subsection{Full mean-field competition model}

\begin{equation}
\frac{\partial\dot{s}_{i}^{P}}{\partial s_{i}^{P}}\equiv J_{ii}^{P}=\alpha_{i}^{P}-2\beta s_{i}^{P}-\beta_{0}\sum_{j\neq i}^{N^{P}} s_{j}^{P}+\gamma_{0}\frac{M_{i}^{P}}{1+h\gamma_{0}M_{i}^{P}},\; i \in P.
\label{eq:JP_ii_fullMF}
\end{equation}

\begin{equation}
\frac{\partial\dot{s}_{i}^{P}}{\partial s_{j}^{P}}\equiv J_{ij}^{P}=-s_i^P\beta_{0},\; i,j \in P.
\label{eq:JP_ij_fullMF}
\end{equation}

\begin{equation}
\frac{\partial\dot{s}_{i}}{\partial s_{k}^{A}}\equiv J_{ik}^{PA}=s_i^P\gamma_{0}\frac{K_{ik}}{\left(1+h\gamma_{0}M_{i}^{P}\right)^{2}},\; i\in P, k\in A.
\label{eq:JPA_ik_fullMF}
\end{equation}

\subsection{Soft mean-field competition model}

\begin{equation}
J_{ii}^{P}=\alpha_{i}^{P}-2\beta s_{i}^{P}-\beta_{0}\sum_{j\neq i}^{N^{P}}A_{ij}^{P}s_{j}^{P}+\gamma_{0}\frac{M_{i}^{P}}{1+h\gamma_{0}M_{i}^{P}},\; i \in P.
\label{eq:JP_ii_softMF}
\end{equation}

\begin{equation}
J_{ij}^{P}=-s_i^P\beta_{0}A_{ij}^{P},\; i,j \in P.
\label{eq:JP_ij_softMF}
\end{equation}

\begin{equation}
J_{ik}^{PA}=s_i^P\gamma_{0}\frac{K_{ik}}{\left(1+h\gamma_{0}M_{i}^{P}\right)^{2}},\; i \in P, k \in A. 
\label{eq:JPA_ik_softMF}
\end{equation}

\subsection{Weighted competition model}

\begin{equation}
J_{ii}^{P}=\alpha_{i}^{P}-2\beta s_{i}^{P}-\beta_{0}\frac{\sum_{j\in P,i\neq j}W_{ij}^{P}s_{j}^{P}}{M_{i}^{P}}+\gamma_{0}\frac{M_{i}^{P}}{1+h\gamma_{0}M_{i}^{P}},\; i \in P.
\label{eq:JP_ii_weighted}
\end{equation}

\begin{equation}
J_{ij}^{P}=-\beta_{0}\frac{s_{i}^{P}}{M_{i}^{P}}W_{ij}^{P},\; i,j \in P.
\label{eq:JP_ij_weighted}
\end{equation}

\begin{equation}
J_{ik}^{PA}=-s_{i}^{P}\beta_{0}K_{ik}\left[\frac{M_{k}^{A}}{M_{i}^{P}}-\frac{s_{i}^{P}}{M_{i}^{P}}-\frac{\sum_{j\in P,i\neq j}W_{ij}^{P}s_{j}^{P}}{\left(M_{i}^{P}\right)^{2}}\right]+s_i^P\gamma_{0}\frac{K_{ik}}{\left(1+h\gamma_{0}M_{i}^{P}\right)^{2}},\; i \in P, k \in A. 
\label{eq:JPA_ik_weighted}
\end{equation}

\end{document}